\documentclass[aps,amsfonts,amsmath,prd,preprint,nofootinbib]{revtex4}
\newcommand{\beq}{\begin{equation}}
\newcommand{\eeq}{\end{equation}}

\usepackage{epsfig,bbm,cancel,xcolor}
\usepackage[breaklinks=true]{hyperref}

\begin{document}

\title{Higher-dimensional black holes with Dirac-Born-Infeld (DBI) global defects}

\author{Handhika S. Ramadhan$^1$}
\email{hramad@ui.ac.id}
\author{Ilham Prasetyo$^{1,2}$}
\email{ilham.prasetyo@sci.ui.ac.id}
\author{Aulia M.~Kusuma$^1$}
\email{aulia.martha@ui.ac.id}
\affiliation{$^1$Departemen Fisika, FMIPA, Universitas Indonesia, Depok 16424, Indonesia.\\
$^2$Research Center for Physics, Indonesian Institute of Sciences (LIPI), Kompleks PUSPIPTEK Serpong, Tangerang 15310, Indonesia.}

\def\changenote#1{\footnote{\bf #1}}

\begin{abstract}
It is well-known that the exact solution of non-linear $\sigma$ model coupled to gravity can be perceived as an exterior  gravitational field of a global monopole. Here we study Einstein's equations coupled to a non-linear $\sigma$ model with Dirac-Born-Infeld (DBI) kinetic term in $D$ dimensions. The solution describes a metric around a DBI global defects. When the core is smaller than its Schwarzschild radius it can be interpreted as a black hole having DBI scalar hair with deficit conical angle. The solutions exist for all $D$, but they can be expressed as polynomial functions in $r$ only when $D$ is even. We give conditions for the mass $M$ and the scalar charge $\eta$ in the extremal case. We also investigate the thermodynamic properties of the black holes in canonical ensemble. The monopole alter the stability differently in each dimensions. As the charge increases the black hole radiates more, in contrast to its counterpart with ordinary global defects where the Hawking temperature is minimum for critical $\eta$. This behavior can also be observed for variation of DBI coupling, $\beta$. As it gets stronger ($\beta\ll1$) the temperature increases. By studying the heat capacity we can infer that there is no phase transition in asymptotically-flat spacetime. The AdS black holes, on the other hand, undergo a first-ordered phase transition in the Hawking-Page type. The increase of the DBI coupling renders the phase transition happen for larger radius. 

\end{abstract}

\maketitle
\thispagestyle{empty}
\setcounter{page}{1}

\section{Introduction}

General relativity coupled to $\sigma$-model has long been studied in various environments, for example: in generating spontaneous compactification of extra dimensions~\cite{GellMann:1984sj, GellMann:1984mu, GellMann:1985if}. From a more phenomenological cosmology point of view, they can model exact solutions of gravitational field outside a global monopole coming from the spontaneous breaking of non-Abelian (say, of an $SO(3)$) global symmetry. Historically, it was Barriola and vilenkin who first studied spacetime around a global monopole by solving the coupled Einstein-Higgs equations~\cite{Barriola:1989hx}. They obtained approximate solutions showing the exterior metric which is Minkowski-like, albeit non-flat, but with deficit solid angle $\Delta=8\pi G\eta^2$. The corresponding black hole solutions was studied in~\cite{Dadhich:1997mh}, where it describes black hole with scalar (global) hair. This can equivalently be interpreted as a black hole eating up a global monopole. Since monopole is a non-contractible defect, its vacuum manifold can be perceived as the exact solution of the non-linear $\sigma$-model (see, for example,~\cite{vilenkinshellard}). To the best of our knowledge, such approach was first used by Olasagasti and Vilenkin~\cite{Olasagasti:2000gx} when studying global defects in braneworld scenario. Later, such technique was emphasized in~\cite{Tan:2017egu, Prasetyo:2017rij} to derive the (higher-dimensional) black holes carrying global scalar hair. In this method, by choosing an appropriate relation between the spacetime and the field (or the internal) metrics, the Barriola-Vilenkin (BV) metric, which is an approximate solution, can be viewed as an exact solution of gravitating non-linear $\sigma$ model.

In recent years, interests in non-canonical global monopole have been steadily growing. In part, this is caused by the study of $k$-defects~\cite{Babichev:2006cy, Babichev:2007tn} that later gained applications in the cosmological context~\cite{ArmendarizPicon:2000dh, ArmendarizPicon:1999rj, ArkaniHamed:2003uy}. The gravitational fields of $k$-monopole, in particular in the form of: (i) power-law, and (ii) Dirac-Born-Infeld (DBI), have been studied in~\cite{Jin:2007fz, Liu:2009eh}. In~\cite{Prasetyo:2015bga} two of us studied the black hole solutions of such non-canonical defects in $4d$ (A)dS space. It is found that for the power-law $k$-monopole, the corresponding black hole resembles that of Reissner-Nordstrom; having at most three horizons for the dS background. On the other hand, the non-canonicality of the DBI does not generate any additional horizons apart from that of Schwarzschild's; {\it i.e.,} the black hole with DBI global monopole is essentially Schwarszschild having deficit solid angle. The deficit angle $\Delta$ is not affected whatsoever by the non-canonical nature of the monopole. We then pursued further study specifically on the case of power-law $k$-monopole in higher dimensions. Using the language of higher-dimensional charged black holes (for example, see~\cite{Cardoso:2004uz}) we can classify the corresponding extremal states in $4d$ as cold, ultracold, and Nariai black holes~\cite{Prasetyo:2017rij}. The nature of our solutions is such that these three extremal states only exist in four dimensions, not present in\footnote{For $D=5$ our metric breaks down. This happens due to the particular choice of our $\sigma$ model. Different model might (or might not) give different singular dimension.} $D>5$.

In this paper, we apply similar analysis to the case of DBI global defects. We present general exact solutions of higher-dimensional (A)dS black holes with (DBI) global hair. We analyze the extremal cases in each dimension. We also investigate its thermodynamical properties as well as its possible phase transition phenomena. In order to achieve that, this work is organized as follows. In the Section~\ref{sec:EDBI} we consider a toy model of $O(D-1)$ $\sigma$-model with DBI kinetic term in $(D+1)$ dimensions coupled to gravity. Taking the hedgehog ansatz for the $\sigma$-model, in Section~\ref{sec:BHsolutions} we present the exact solutions that can be interpreted as exterior metric of a non-canonical global defects. As in the case with other black holes, our solutions also radiates and behaves like a thermodynamical object. Our investigation reveals distinctive features of thermodynamical properties that differs from their canonical case. In order to show such genuine results, we first review the thermodynamical properties of BV black hole in Section~\ref{sec:thermoreview}. Later in Section ~\ref{sec:thermodbi} we analyze the thermodynamics of our DBI solutions. Section~\ref{sec:factor} is devoted to the factorized solution of our toy model. Finally, we give conclusions in Section~\ref{sec:conclusion}.

\section{Einstein-DBI sigma model theory}\label{sec:EDBI}

As in~\cite{Prasetyo:2017rij}, we consider a toy model described by the following action in $D$-dimensional spacetime
\begin{equation}
\mathcal{S}=\int d^{D}x \; \sqrt{|g|} \;
\left( {R-2\Lambda \over 16\pi G} 
+\mathcal{K}(\mathcal{X})-{\lambda\over 4} \left( \vec{\Phi}^2-\eta^2 \right)
\right). 
\end{equation}
Here $\mathcal{K}(\mathcal{X})$ is a functional of $\mathcal{X}\equiv -(1/2)\partial_M\vec{\Phi}\partial^M\vec{\Phi}$, $\Lambda$ and $G$ are the corresponding $D$-dimensional cosmological and Newton's constants, respectively. From the point of view of $\sigma$-model, the last term is not a potential but a constraint enforced to be satisfied by the fields $\vec{\Phi}$, while the constant $\lambda$ acts as its Lagrange multiplier. In order to avoid ``zero-kinetic problem" on the one hand and can reduce to the canonical case on the other hand, the functional  $\mathcal{K}(\mathcal{X})$ should satisfy~\cite{Babichev:2006cy, Jin:2007fz}:
\begin{eqnarray}
\mathcal{K}(\mathcal{X})=\begin{cases}
-\mathcal{X},\ \ \  |\mathcal{X}|\ll1,\\
-\mathcal{X}^{\alpha},\ \ \ |\mathcal{X}|\gg1,
\end{cases}
\label{chi}
\end{eqnarray}
with $\alpha$ a positive constant. In this paper we choose to work with
\begin{equation}
\mathcal{K}(\mathcal{X})\equiv \beta^2\left(1-\sqrt{1+{2\mathcal{X}\over \beta^2}}\right).
\label{eq:kdbi}
\end{equation}
This choice is, in part, motivated by the action for $D3$-brane on warped background~\cite{Alishahiha:2004eh}. 

In general the scalar field $\vec{\Phi}$ enjoys an $O(d)$ symmetry. Due to the constraint term, it is spontaneously broken to $O(d-1)$. This breaking forces $\vec{\Phi}$ to stay on its vacuum manifold $\mathcal{M}$, defined by $\vec{\Phi}^2=\eta^2$, topologically an $S^{d-1}$. The field can then be perceived as having internal coordinates, $\vec{\Phi}=\vec{\Phi}\left(\phi^i\right)$, $i=1, 2,\cdots, d-1$. The effective action is 
\begin{equation}
\mathcal{S}=\int d^{D}x~ \sqrt{|g|} \left[
{R-2\Lambda\over 16\pi G}+
\beta^2\left(
1-\sqrt{1-{\eta^2 h_{ij}\partial_M \phi^i \partial^M \phi^j\over \beta^2}}
\right)\right].\label{eq:001}
\end{equation}
The vector is then defined by $\phi^i$ with its inner space metric $h_{ij}=h_{ij}(\phi^k)$.

Varying the action~\eqref{eq:001} we obtain equation of motions for the scalar field and the energy-momentum tensor 
\begin{equation}
{2\over\sqrt{|g|}}\partial_M\left(
{\sqrt{|g|} h_{ij} \partial^M\phi^j \over \sqrt{1-{\eta^2h_{ij}\partial_M \phi^i \partial^M \phi^j\over \beta^2}}}
\right)={\partial_M\phi^p\partial^M\phi^q\over \sqrt{1-{\eta^2h_{ab}\partial_M \phi^a \partial^M \phi^b\over \beta^2}}}{\partial h_{pq}\over\partial\phi^i},\label{eq:002}
\end{equation}
\begin{equation}
T^M_N=\delta^M_N\left[
{\Lambda\over 8\pi G}-\beta^2\left(
1-\sqrt{1-{\eta^2h_{ij}\partial_M \phi^i \partial^M \phi^j\over \beta^2}}
\right)
\right]+{\eta^2h_{ij}\partial^M\phi^i \partial_N\phi^j \over\sqrt{1-{\eta^2h_{ij}\partial_M \phi^i \partial^M \phi^j\over \beta^2}}}.\label{eq:003}
\end{equation}
Here we use a spherically-symmetric metric ansatz
\begin{equation}
ds^2=g_{MN} dx^M dx^N=A^2(r)dt^2 -B^2(r)dr^2-C^2(r)d\Omega^2_{D-2},
\end{equation}
with $d\Omega^2_{D-2}=\gamma_{ij}(\theta^k)d\theta^i d\theta^j$. It has a unit $(D-2)$-sphere parametrized by angular coordinates $\theta^1,\theta^2,...,\theta^{D-2}$ and $D<\infty$. The Ricci tensor components are thus
\begin{eqnarray}
R^0_0&=&B^{-2}\left[
{A''\over A}-{A'B'\over AB} +(D-2) {A'C'\over AC}
\right],\\
R^r_r&=&B^{-2}\left[
{A''\over A} + (D-2) {C''\over C} -{B'\over B}\left\{
{A'\over A} +(D-2) {C'\over C}
\right\}\right],\\
R^\theta_\theta&=&B^{-2}\left[
{C''\over C}+{C'\over C}\left\{
{A'\over A}-{B'\over B}+(D-3){C'\over C} \right\}\right]-{(D-3)\over C^2}.
\end{eqnarray}

As was employed in Ref.~\cite{Prasetyo:2017rij} (and was also independently pointed out in~\cite{Tan:2017egu}) our toy model can describe exterior solution of an $O(d-1)$ global defects should we pick the appropriate ans\"{a}tz. The simplest one we can take is the hedgehog
\begin{equation}
\phi^i=\phi^i(\theta^j)=\theta^i,
\end{equation}
where $i=1, \cdots, D-2$. This choice means that the number of degrees of freedom of the internal space is equal the angular degrees of freedom of the coordinate space. It is not difficult to see that this ans\"{a}tz satisfies Eq.\eqref{eq:002} provided we assume
\begin{equation}
h_{ij}(\phi^k)=-{1\over C^2(r)}g_{ij}(r,\theta^k)=\gamma_{ij}(\theta^k).
\end{equation}
The energy-momentum tensor now becomes
\begin{eqnarray}
8\pi G T^0_0&=&{\Lambda}-8\pi G\beta^2\left(
1-\sqrt{1+{(D-2)\eta^2\over \beta^2C^2}}
\right)=8\pi G T^r_r,\\
8\pi G T^\theta_\theta&=&8\pi G T^r_r-{8\pi G\eta^2/C^2\over\sqrt{1+{(D-2)\eta^2\over \beta^2C^2}}}.
\end{eqnarray}


\section{A black hole eating up a global defect}\label{sec:BHsolutions}

In this section we wish to study black hole solutions of our model. This is done by choosing $C(r)=r$. The Einstein equations $R^A_{B}=8\pi G(T^A_B-\delta^A_B {T\over (D-2)})$ become
\begin{eqnarray}
&& R^0_0={1\over B^2}\left[ {A''\over A} - {A'B'\over AB} + (D-2) {A'\over rA}
\right]=-{16\pi G\over D-2} T^0_0 + {8\pi G\eta^2/r^2\over \sqrt{1+{(D-2)\eta^2\over \beta^2 r^2}}},\\
&& R^r_r={1\over B^2}\left[ {A''\over A} - {A'B'\over AB} - (D-2) {B'\over rB}
\right]=-{16\pi G\over D-2} T^0_0 + {8\pi G\eta^2/r^2\over \sqrt{1+{(D-2)\eta^2\over \beta^2 r^2}}},\\
&& R^\theta_\theta = {1\over B^2}\left[ {A'\over rA} - {B'\over rB} + {D-3\over r^2}
\right] -{D-3\over r^2} 
=-{16\pi G\over D-2} T^0_0.
\end{eqnarray}
From $R^0_0-R^r_r$ we conclude $A=B^{-1}.$ Substituting it back into $R^\theta_\theta$ gives us
\begin{equation}
 \label{eq:gensol}
R^\theta_\theta = {1\over r^{D-2}}\left( {r^{D-3}\over B^2} \right)'- {D-3\over r^2}=-{2\over D-2}\left[ \Lambda -8\pi G\beta^2 \left(1-\sqrt{1+{(D-2)\eta^2\over \beta^2 r^2}} \right)\right].
\end{equation}

\subsection{Black hole in even dimensions}

Eq.~\eqref{eq:gensol} in general can easily be integrated, and we obtain (with $M$ as a constant of integration)
\begin{eqnarray}
B^{-2}&=&\frac{D^2-16 \pi  \beta ^2 G r^2 \, _2F_1\left(-\frac{1}{2},\frac{1-D}{2};\frac{3-D}{2};-\frac{(D-2) \eta ^2}{r^2 \beta ^2}\right)-3 D+16 \pi  \beta ^2 G r^2-2 \Lambda  r^2+2}{(D-2) (D-1)}
-{2G M\over r^{D-3}}\nonumber\\
&=&1-{2G M\over r^{D-3}}-{2 \Lambda  r^2 \over (D-2) (D-1)}-{16 \pi  \beta ^2 G r^2\left(\, _2F_1\left(-\frac{1}{2},\frac{1-D}{2};\frac{3-D}{2};-\frac{(D-2) \eta ^2}{r^2 \beta ^2}\right)-1\right) \over (D-2) (D-1)}, \label{eq:000}
\end{eqnarray}
with $\, _2F_1\left(a,b;c;d\right)$ is the hypergeometric function~\cite{abramowitz}. This is the exact solution of the gravitational field outside a higher-dimensional DBI global defects, and the constant $M$ can be interpreted as the black hole mass. When the mass is greater than its typical defect core, $M\gg\delta$, the solution describes a black hole with a global charge.

To investigate the singularity we calculate the Kretchmann scalar
\begin{eqnarray}
R^{ABCD}R_{ABCD}&=&{32 (D-3) K_1^2\over (D-2) (D-1)^2 \ r^{2 D} }+{4 \beta ^2 K_2^2\over (D-2)^2 (D-1)^2\ r^{2 (D+1)}\left((D-2) \eta ^2+\beta ^2 r^2\right)}\nonumber\\
&&+\frac{2 (D-2) K_3^2}{r^2},
\end{eqnarray}
with
\begin{equation}
K_1\equiv
G \left(\left(D^2-3 D+2\right) M r-8 \pi  \beta ^2 r^D\right)+8 \pi  \beta ^2 G r^D \, _2F_1\left(-\frac{1}{2},\frac{1-D}{2};\frac{3-D}{2};-\frac{(D-2) \eta ^2}{r^2 \beta ^2}\right)+\Lambda  r^D,\nonumber\\
\end{equation}
\begin{eqnarray}
K_2&\equiv&D^4 G M r^3 \sqrt{\frac{(D-2) \eta ^2+\beta ^2 r^2}{\beta ^2 r^2}}-8 D^3 G M r^3 \sqrt{\frac{(D-2) \eta ^2+\beta ^2 r^2}{\beta ^2 r^2}}-8 \pi  D^3 \eta ^2 G r^D \nonumber\\
&&+8 \pi  \beta ^2 \left(D^2-5 D+6\right) G r^{D+2} \sqrt{\frac{(D-2) \eta ^2+\beta ^2 r^2}{\beta ^2 r^2}} \, _2F_1\left(-\frac{1}{2},\frac{1-D}{2};\frac{3-D}{2};-\frac{(D-2) \eta ^2}{r^2 \beta ^2}\right)\nonumber\\
&&+23 D^2 G M r^3 \sqrt{\frac{(D-2) \eta ^2+\beta ^2 r^2}{\beta ^2 r^2}}-8 \pi  \beta ^2 D^2 G r^{D+2}+48 \pi  D^2 \eta ^2 G r^D\nonumber\\
&&-28 D G M r^3 \sqrt{\frac{(D-2) \eta ^2+\beta ^2 r^2}{\beta ^2 r^2}}+12 G M r^3 \sqrt{\frac{(D-2) \eta ^2+\beta ^2 r^2}{\beta ^2 r^2}}+40 \pi  \beta ^2 D G r^{d+2}\nonumber\\
&&-32 \pi  \beta ^2 G r^{D+2}-88 \pi  D \eta ^2 G r^D+48 \pi  \eta ^2 G r^D-16 \pi  \beta ^2 G r^{D+2} \sqrt{\frac{(D-2) \eta ^2+\beta ^2 r^2}{\beta ^2 r^2}}\nonumber\\
&&+2 \Lambda  r^{D+2} \sqrt{\frac{(D-2) \eta ^2+\beta ^2 r^2}{\beta ^2 r^2}},\nonumber
\end{eqnarray}
\begin{equation}
K_3\equiv2 (D-3) G M r^{2-D}-\frac{4 r K_4}{(D-2) (D-1)},\nonumber
\end{equation}
\begin{eqnarray}
K_4&\equiv&-4 \pi  \beta ^2 (D-3) G \, _2F_1\left(-\frac{1}{2},\frac{1-D}{2};\frac{3-D}{2};-\frac{(D-2) \eta ^2}{r^2 \beta ^2}\right)\nonumber\\
&&+4 \pi  \beta ^2 G \left(D \sqrt{\frac{(D-2) \eta ^2+\beta ^2 r^2}{\beta ^2 r^2}}-\sqrt{\frac{(D-2) \eta ^2+\beta ^2 r^2}{\beta ^2 r^2}}-2\right)+\Lambda.
\end{eqnarray}
It is not difficult to see that there the scalar is divergent at $r=0$. To identify whether this solution describes a black hole or naked singularity, we should check whether there exists horizon(s) or not.

Expanding the hyperbolic function $_2F_1$ of eq. \eqref{eq:000} with respect to $1/r$ we obtain
\begin{eqnarray}
B^{-2}&=&-\frac{7 \eta ^{10} \left(\pi  (D-2)^4 G\right)}{16 \left(\beta ^8 (D-11) r^8\right)}+\frac{5 \pi  (D-2)^3 \eta ^8 G}{8 \beta ^6 (D-9) r^6}-\frac{\pi (D-2)^2 \eta ^6 G}{\beta ^4 (D-7) r^4}+\frac{2 \pi  (D-2) \eta ^4 G}{\beta ^2 (D-5) r^2}+\mathcal{O}(r^{-9})\nonumber\\
&&+\left(1-\frac{2 \Lambda  r^2}{D^2-3 D+2}\right)-\frac{8 \eta ^2 (\pi  G)}{D-3}-{2G M\over r^{D-3}}\nonumber\\
&=&1-\frac{8 \eta ^2 \pi  G}{D-3}-\frac{2 \Lambda  r^2}{(D-2) (D-1)}-{2G M\over r^{D-3}} +\frac{2 \pi  (D-2) \eta ^4 G}{\beta ^2 (D-5) r^2}
\nonumber\\
&&-\frac{7 \eta ^{10} \left(\pi  (D-2)^4 G\right)}{16 \left(\beta ^8 (D-11) r^8\right)}+\frac{5 \pi  (D-2)^3 \eta ^8 G}{8 \beta ^6 (D-9) r^6}-\frac{\pi  (D-2)^2 \eta ^6 G}{\beta ^4 (D-7) r^4}+\mathcal{O}(r^{-9}).\label{eq:expandedtangherlini}
\end{eqnarray}
This gives us an impression that the solution exists only for even dimensions higher than three. For $\beta\gg1$ Eq.~\eqref{eq:expandedtangherlini} simply reduces to the ordinary Tangherlini black hole (\cite{Tangherlini:1963bw}) with global defects,
\begin{equation}
B^{-2}=1-\frac{8 \eta ^2 \pi  G}{D-3}-\frac{2 \Lambda  r^2}{(D-2) (D-1)}-{2G M\over r^{D-3}}.\label{eq:tangherlini}
\end{equation}
By rescaling with 
$t=\bar{t}(1-\Delta)^{-1/2}$, $r=\bar{r}(1-\Delta)^{1/2}$ where
\begin{equation}
\Delta\equiv\frac{8 \eta ^2 \pi  G}{D-3}\label{eq:delta}
\end{equation}
is the solid deficit angle, we obtain the following metric solution
\begin{equation}
ds^2=\bar{B}^{-2}d\bar{t}^2-{d\bar{r}^2\over \bar{B}^{-2}}-(1-\Delta)\bar{r}^2d\Omega^2_{(D-2)},\label{eq:metricdelta}
\end{equation}
with


\begin{eqnarray} 
\bar{B}^{-2}&=&\left\{1-\Delta-\frac{2 \Lambda  (\bar{r}(1-\Delta)^{1/2})^2}{(D-2) (D-1)}-{2G M\over (\bar{r}(1-\Delta)^{1/2})^{D-3}} +\frac{2 \pi  (D-2) \eta ^4 G}{\beta ^2 (D-5) (\bar{r}(1-\Delta)^{1/2})^2}\right.
\nonumber\\
&&-\frac{7 \eta ^{10} \left(\pi  (D-2)^4 G\right)}{16 \left(\beta ^8 (D-11) (\bar{r}(1-\Delta)^{1/2})^8\right)}+\frac{5 \pi  (D-2)^3 \eta ^8 G}{8 \beta ^6 (D-9) (\bar{r}(1-\Delta)^{1/2})^6}\nonumber\\
&&-\left.\frac{\pi  (D-2)^2 \eta ^6 G}{\beta ^4 (D-7) (\bar{r}(1-\Delta)^{1/2})^4}+\mathcal{O}((\bar{r}(1-\Delta)^{1/2})^{-9})\right\}/(1-\Delta) \nonumber\\
&=&1-\frac{2 \Lambda  \bar{r}^2}{(D-2) (D-1)}-{2\bar{G} \bar{M}\over \bar{r}^{D-3}} +\frac{2 \pi  (D-2) {\eta} ^4 \bar{G}}{\bar{\beta} ^2 (D-5) \bar{r}^2}
\nonumber\\
&&-\frac{\pi  (D-2)^2 {\eta} ^6 \bar{G}}{\bar{\beta} ^4 (D-7) \bar{r}^4}+\frac{5 \pi  (D-2)^3 {\eta} ^8 \bar{G}}{8 \bar{\beta} ^6 (D-9) \bar{r}^6}-\frac{7 {\eta} ^{10} \left(\pi  (D-2)^4 \bar{G}\right)}{16 \left(\bar{\beta} ^8 (D-11) \bar{r}^8\right)}+\mathcal{O}(\bar{r}^{-9}),\label{eq:expandbars}
\end{eqnarray}
by rescaling some constants $\bar{M}=M(1-\Delta)^{(3-D)/2}, \bar{G}=G(1-\Delta)^{-1}$ and $\bar{\beta}=\beta(1-\Delta)^{1/2}$.


Looking at the terms in~\eqref{eq:expandbars}, we can see that they can be rewritten as
\begin{eqnarray}
\bar{B}^{-2}\equiv f_D(r)&=&1-{2\bar{G} \bar{M}\over \bar{r}^{D-3}}+\frac{8 {\eta} ^2 \pi  \bar{G}}{D-3}-{2 \Lambda \bar{r}^2 \over (D-2) (D-1)}\nonumber\\
&&-{16 \pi  \bar{\beta} ^2 \bar{G} \bar{r}^2\left(\, _2F_1\left(-\frac{1}{2},\frac{1-D}{2};\frac{3-D}{2};-\frac{(D-2) {\eta} ^2}{\bar{r}^2 \bar{\beta} ^2}\right)-1\right) \over (D-2) (D-1)}.
\label{eq:004}
\end{eqnarray} 
We then remove the bars for convenience. To guarantee that the spacetime is free of conical singularity, the symmetry breaking scale is constrained to be $\eta<\eta_{crit}\equiv\sqrt{D-3\over8\pi G}$. The metric~\eqref{eq:metricdelta} with $f(r)=B^{-2}(r)$ given by~\eqref{eq:004} describes higher-dimensional black hole possessing global scalar hair. This may sound violating the well-known no-hair conjecture, but recall that the space around this black hole is not asymptotically-flat, but suffers from deficit solid angle. Thus, the conjecture does not apply here. 

The horizon(s) are then the roots of 
\begin{equation}
f_D(r_\pm)=0,
\end{equation}
where $r_{\pm}$ are the horizon(s) radius. It appears that for  $\Lambda>0$ we can have two horizons, while for $\Lambda\leq0$ there exists only one horizon. The mass of the black hole $M$ can then be determined (see, for example,~\cite{Cai:2004eh, Myung:2008eb}) and is given by
\begin{eqnarray}
M_D&=&\frac{r_+^{D-3}}{2 G}+\frac{8 \pi  b^2 r_+^{D-1}}{(D-2) (D-1)}-\frac{\Lambda 
   r_+^{D-1}}{G (D-2) (D-1)}+\frac{4 \pi  \eta ^2 r_+^{D-3}}{D-3}\nonumber\\
   &&-\frac{8 \pi  b^2 r_+^{D-1} \, _2F_1\left(-\frac{1}{2},\frac{1-D}{2};\frac{3-D}{2};-\frac{(D-2) {\eta} ^2}{r_+^2 \bar{\beta} ^2}\right)}{(D-2) (D-1)},
   \label{eq:massex}
\end{eqnarray} 
where $r_+$ is the radius of the outer horizon. In the next subsections we shall discuss in detail the (A)dS black hole with DBI global monopole in various dimensions along with its corresponding extremal states.


\subsubsection{$D=4$ case}

In $D=4$, the metric~\eqref{eq:004} becomes
\begin{eqnarray}
f_4(r)&=&1+8 \pi  G \eta ^2-\frac{2 G M}{r}-\frac{r^2 \Lambda }{3}-\frac{8}{3} \pi   \beta^2 G r^2 \left(\left(\frac{2 \eta ^2}{ \beta^2 r^2}+1\right)^{3/2}-1\right).
\label{eq:exact008}
\end{eqnarray}
This case has been discussed in~\cite{Prasetyo:2015bga}, where the metric solution obtained is the same as Eq.~\eqref{eq:exact008} above.
The black hole mass~\eqref{eq:massex} is given by
\begin{equation}
M_4=\frac{r_+ \left(1+8 \pi  G \eta
   ^2-\frac{r_+^2 \Lambda }{3}-\frac{8}{3} \pi  b^2 G r_+^2 \left(\left(\frac{2 \eta ^2}{b^2 r_+^2}+1\right)^{3/2}-1\right)\right)}{2 G}.
\end{equation}
To understand the metric better, let us expand it in terms of\footnote{Or in terms of $\eta^2$, if you like.} $r\beta$
\begin{equation}
f_4(r)=1-{\Lambda r^2 \over 3}-{2G M\over r} -\frac{4 \pi {\eta} ^4 G}{\beta ^2 r^2}+\frac{4\pi {\eta} ^6 G}{3\beta ^4 r^4}-\frac{\pi {\eta} ^8 G}{ \beta ^6r^6}+\frac{{\eta} ^{10} \pi G}{\beta ^8 r^8}+\mathcal{O}((r \beta)^{-9}).
\label{eq:008}
\end{equation}
In this expanded form it is manifest that, unlike the black hole with power-law global monopole~\cite{Prasetyo:2017rij} whose existence of the scalar charge resembles Reissner-Nordstrom-de Sitter (RNdS) black hole, in this metric solution the scalar charge term has negative sign. This makes the black hole have one less horizon(s) than its RNdS counterpart. 

For $\Lambda=0$ clearly there is only one horizon. This can be seen by looking at~\eqref{eq:008} and keeping only terms up to $(\eta^2/r\beta)^2$. The approximate horizon lies on
\begin{equation}
r_\pm\approx{GM}\pm\sqrt{({GM})^2+4 \pi {\eta} ^4 G\beta^{-2}}.
\end{equation}
Thus, for any positive values of $G, M, \beta$ and $\eta$ only $r_+$ is positive.


For $\Lambda\neq0$, there can be at most two horizons. The extremal can be determined by setting (see~\cite{Myung:2008eb}):
\begin{equation}
f_D(r_e)=0,\ \ \ \ {df_D(r)\over dr}\bigg|_{r=r_e}=0,\label{eq:extremecondition}
\end{equation}
where $r_e$ is the extremal horizon. These conditions lead to the following algebraic equation for $r_e$:
\begin{equation}
1+8 \pi  G \eta ^2+8 \pi  b^2 G r_e^2-8 \pi  b^2 G r_e^2 \sqrt{\frac{2 \eta ^2}{b^2 r_e^2}+1}-r_e^2 \Lambda=0,
\end{equation}
whose solutions are
\begin{equation}
r_e^2=\frac{-\Lambda+8 \pi  \left(+b^2 G-G \eta ^2 \Lambda\pm\sqrt{b^2 G^2 \left(b^2+16 \pi  G \eta ^4 \Lambda +2 \eta ^2 \Lambda \right)}\right)}{\Lambda  \left(16 \pi 
   b^2 G-\Lambda \right)}.\label{eq:re4}
\end{equation}
It can be shown that unless $\Lambda>0$, $r_e^2<0$ . The extremal mass, $M_{ext}$, is the mass~\eqref{eq:massex} with $r_+=r_e$.

Even though the method above is exact, it is instructive to consider the expanded solution~\eqref{eq:008} by keeping only terms up to $\eta^4$, since this form shall give us information about the relation between the mass $M$ and the global charge $\eta$: 
\begin{equation}
f_4(r)\simeq1-{\Lambda r^2 \over 3}-{2G M\over r}-\frac{4 \pi {\eta} ^4 G}{\beta ^2 r^2}.\label{eq:005}
\end{equation}
This can be re-expressed as
\begin{equation}
f_4(r)\simeq\left(1-{\rho \over r}\right)^2 \left(1-{\Lambda\over 3}\left(r^2+a+br\right)\right),
\end{equation}
where $\rho$, $a$ and $b$ are constants. After a short calculation we obtain \[b=2\rho,~ a=3\rho^2,\]
where
\begin{eqnarray}
M&\simeq&{\rho\over G}\left(1-{2\Lambda\over 3}\rho^2\right),\\
{\eta^4\over\beta^2}&\simeq&{\rho^2\over 4\pi G}\left(-1+\Lambda\rho^2\right).
\end{eqnarray}
The above expression enables us to extract constraint on the cosmological constant. It must have a value between ${1\over \rho^2}\leq\Lambda<{3\over 2\rho^2}$.
At 
\begin{equation}
\Lambda\simeq{1\over \rho^2}
\label{eq:006}
\end{equation}
the black hole becomes extremal. It agrees approximately with~\eqref{eq:re4}. This extremal condition is satisfied when
\begin{equation}
{4\pi G\eta^4\over 3\beta^2}\simeq0.
\label{eq:007}
\end{equation}
Note that since~\eqref{eq:005} is an approximate solution, then condition above implies that $\eta^2/\beta$ is very small. At this value, we have
\begin{equation}
\Lambda\simeq{1\over (3GM)^2}.
\end{equation}
This approach is only valid for weak coupling  case or when $\eta$ is much smaller than $1/\sqrt{8\pi G}$.

\subsubsection{$D=6$ case}

The metric in this dimension is
\begin{equation}
f_6(r)=1+\frac{8}{3} \pi  G \eta ^2-\frac{2 GM}{r^3}-\frac{\Lambda r^2}{10}-\frac{4}{5} \pi  b^2 G r^2 \left(\frac{\left(3 b^2 r^2-8 \eta ^2\right) \left(\frac{4 \eta ^2}{b^2 r^2}+1\right)^{3/2}}{3 b^2 r^2}-1\right),
\end{equation}
or, in its expanded version, 
\begin{equation}
f_6(r)=1-{2GM\over r^{3}}-\frac{ \Lambda  r^2}{10}+\frac{8 \pi {\eta} ^4G}{\beta ^2 r^2} +\frac{ 16 \pi  {\eta} ^6 G}{\beta ^4 r^4}-\frac{40 \pi {\eta} ^8 G}{3\beta^6  r^6}+\frac{112 {\eta} ^{10} \pi G}{5\beta^8 r^8}+\mathcal{O}(r^{-9}).\label{eq:expand6}
\end{equation}
The mass is given by
\begin{equation}
M_6=\frac{r_+^3 \left(1+\frac{8}{3} \pi  G \eta ^2-\frac{r_+^2 \Lambda }{10}+\frac{4}{5} \pi  \beta^2 G r_+^2 \left(1-\frac{\left(3 \beta^2 r_+^2-8 \eta ^2\right) \left(\frac{4 \eta ^2}{\beta^2 r_+^2}+1\right)^{3/2}}{3 \beta^2
   r_+^2}\right)\right)}{2 G}.
\end{equation}
By employing conditions~\eqref{eq:extremecondition}, the extremal horizon $r_e$ is the one that solves
\begin{equation}
\frac{3}{r_e}+\frac{8 \pi  G \eta ^2}{r_e}-\frac{r_e \Lambda }{2}+4 \pi  \beta^2 G r_e-4 \pi  \beta^2 G r_e \sqrt{\frac{4 \eta ^2}{\beta^2 r_e^2}+1}=0.
\end{equation}
It is given by
\begin{equation}
r_e=\sqrt{\frac{-6 \Lambda+16 \pi  \left(3 \beta^2 G-G \eta ^2 \Lambda\pm\sqrt{\beta^2 G^2 \left(9 \beta^2+16 \pi  G \eta ^4 \Lambda +6 \eta ^2 \Lambda \right)}\right)}{\Lambda 
   \left(16 \pi  \beta^2 G-\Lambda \right)}}.\label{eq:re6}
\end{equation}


Looking at the form~\eqref{eq:expand6} (assuming $\eta<<\eta_{crit}$) we have, by truncating it up to $\eta^6$,
\begin{equation}
f_6(r)\simeq1-{2GM\over r^{3}} -\frac{ \Lambda r^2}{10}+\frac{8 \pi {\eta} ^4 G}{\beta ^2 r^2}+\frac{ 16 \pi  {\eta} ^6 G}{\beta ^4 r^4}.
\end{equation}
This can be approximated by
\begin{equation}
f_6(r)\simeq
\left(1-{\rho \over r}\right)^2 \left(1-{\Lambda\over 10}\left(r^2+a+br+{c_1\over r}+{c_2\over r^2}\right)\right).
\end{equation}
One obtain
\[a=3\rho^2,~~b=2\rho,~~c_1=4\rho^3-{20\rho\over \Lambda},~~c_2={1\over\rho}{16\pi G\eta^6\over\beta^4},\]
and thus
\begin{eqnarray}
2GM &\simeq&2\rho^3\left(-1+{\Lambda\rho^2\over5}\right)-{\Lambda\over5\rho}{16\pi G\eta^6\over\beta^4},\label{pers07}\\
{8\pi G\eta^4\over\beta^2} &\simeq&\rho^2\left(-3+{\Lambda\rho^2\over2}\right)-{\Lambda\over10\rho^2}{16\pi G\eta^6\over\beta^4}.\label{pers07a}
\end{eqnarray}
For small enough $\eta^3/\beta^2$ we can safely approximate  ${\eta^6\over\beta^4}\sim0$, and from condition~\eqref{pers07} the cosmological constant is bounded from below, 
\begin{equation}
\Lambda\gtrsim{5\over\rho^2}.\label{eq:pers07b}
\end{equation}
Since this approximation is valid for small $\eta$, from~\eqref{pers07a} the extremal state happens when $\Lambda\simeq{6\over\rho^2}$, or $\rho\simeq\sqrt{6\over\Lambda}$, which agrees with the approximate solution~\eqref{eq:re6} and is also consistent with~\eqref{eq:pers07b}. At this value, one obtains
\begin{equation}
\Lambda\simeq{6\over (5GM)^{2/3}}.
\end{equation}

\subsubsection{$D=8$ case}

The metric solution in this case becomes
\begin{eqnarray}
f_8(r)&=&1+\frac{8\pi}{5} G \eta ^2-\frac{2 G M}{r^5}-\frac{r^2 \Lambda }{21}-\frac{8 \pi}{21}\beta^2 G r^2 \left(\frac{\left(\frac{6 \eta ^2}{\beta^2 r^2}+1\right)^{3/2} \left(5 \beta^4 r^4-24 \beta^2 r^2 \eta ^2+96 \eta ^4\right)}{5 \beta^4
   r^4}-1\right)\nonumber\\
   &=&1-{2GM\over r^{5}}-\frac{ \Lambda  r^2}{21} +\frac{4 \pi {\eta} ^4 G}{\beta^2 r^2} -\frac{36\pi {\eta} ^6 G}{\beta^4 r^4}-\frac{145 \pi {\eta} ^8 G}{\beta^6 r^6}+\mathcal{O}(r^{-7}),\label{eq:sol8d}
\end{eqnarray}
while the mass is given by
\begin{equation}
M_8=\frac{r_+^5 \left(1+\frac{8}{5} \pi  G \eta ^2-\frac{r_+^2 \Lambda }{21}-\frac{8}{21} \pi  \beta^2 G r_+^2 \left(\frac{\left(\frac{6 \eta ^2}{\beta^2 r_+^2}+1\right)^{3/2} \left(5 \beta^4 r_+^4-24 \beta^2 r_+^2 \eta ^2+96 \eta ^4\right)}{5
   \beta^4 r_+^4}-1\right)\right)}{2 G}.
\end{equation}
To obtain the extremal horizon, we follow the same procedure as before and arrive at
\begin{equation}
-\frac{8}{3} \pi \beta^2 G r_e \sqrt{\frac{6 \eta^2}{\beta^2 r_e^2}+1}+\frac{8}{3} \pi  \beta^2 G r_e+\frac{8 \pi  G \eta ^2}{r_e}-\frac{r_e \Lambda }{3}+\frac{5}{r_e}=0.
\end{equation} 
The solution is
\begin{equation}
r_e=\sqrt{3} \sqrt{\frac{-5
   \Lambda+8 \pi  \left(5 \beta^2 G-G \eta ^2 \Lambda\pm\sqrt{\beta^2 G^2 \left(25\beta^2+2 \eta ^2 \Lambda  \left(8 \pi  G \eta ^2+5\right)\right)}\right)}{\Lambda  \left(16 \pi  \beta^2 G-\Lambda \right)}}.
\end{equation}

Perturbatively, we can approach ~\eqref{eq:sol8d} with
\begin{equation}
f_8(r)\simeq
\left(1-{\rho \over r}\right)^2 \left(1-{\Lambda\over 10}\left(r^2+a+b r+{c_1\over r}+{c_2\over r^2}\right)\right),
\end{equation}
whose coefficients are
\[
a=3 \rho ^2,~~
b = 2 \rho,~~
c_1=\frac{2 \left(2 \Lambda  \rho ^3-21 \rho \right)}{\Lambda },~~
c_2=\frac{5 \Lambda  \rho ^4-21 {k_1}-63 \rho ^2}{\Lambda },
\]
\[
c_3=-\frac{6 \left(-\Lambda  \rho ^5+7 k_1 \rho +14 \rho ^3\right)}{\Lambda },~~
c_4=\frac{7 \left(\Lambda  \rho ^6-9 k_1 \rho ^2+3 k_2-15 \rho ^4\right)}{\Lambda },
\]
and
\[
k_1=\frac{4 \pi {\eta} ^4 \bar{G}}{\bar{\beta} ^2},~~
k_2=\frac{36\pi {\eta} ^6 \bar{G}}{\bar{\beta} ^4},~~
k_3=\frac{145 \pi {\eta} ^8 \bar{G}}{\bar{\beta} ^6}.
\]
After some little algebra we obtain
\begin{eqnarray}
2GM &\simeq& \frac{-2}{21}\left(4 \Lambda  \rho ^7-42 \left(\frac{4 \pi {\eta} ^4 G}{\beta^2}\right) \rho^3+21 \left(\frac{36\pi {\eta} ^6 G}{\beta^4}\right) \rho -63 \rho ^5\right),\label{eq:d8a}\\
\frac{4 \pi {\eta} ^4 G}{\beta^2} &\simeq& \frac{\Lambda  \rho ^8+3 \left(\frac{36\pi {\eta} ^6 G}{\beta^4}\right) \rho ^2-3 \left(\frac{145 \pi {\eta} ^8 G}{\beta^6}\right)-15 \rho ^6}{9 \rho ^4}.\label{eq:d8b}
\end{eqnarray}
Ignoring the term ${\eta^6\over\bar{\beta}^4}$, condition~\eqref{eq:d8b} requires
\begin{equation}
\Lambda\simeq{15\over\rho^2},
\end{equation}
or, equivalently, $\rho\simeq\sqrt{15\over\Lambda}$. It is easy to see that it guarantees condition~\eqref{eq:d8a} to have positive mass. At this value, we have
\begin{equation}
\Lambda\simeq{15\over (7GM)^{2/5}}.
\end{equation}


In principle, the methods employed above can be applied to any arbitrary higher dimension ($D=10, 12,....$).

\subsection{Black hole in odd dimensions}

As stated above, the structure of our theory gives an impression that the solutions only exist in even dimensions, as given by~\eqref{eq:expandedtangherlini}. This impression is misleading. It is due to the mathematical structure of Eq.~\eqref{eq:gensol}. The dependence of the hypergeometric function $\, _2F_1\left(-\frac{1}{2},\frac{1-D}{2};\frac{3-D}{2};-\frac{(D-2) \eta ^2}{r^2 \beta ^2}\right)$ in (its general solution) on the radius always come in the combination of $\beta r$. In this sense condition~\eqref{eq:expandedtangherlini} is valid\footnote{We thank the anonymous referee for pointing this out.} only for large $\beta r$. 

To obtain exact solutions for odd dimensions, we cannot use Eq.\eqref{eq:000}. Instead we should return back to Eq.~\eqref{eq:gensol}. The integral can be solved for any specific $D$. Take, for example, $D=5$. The equation becomes
\begin{equation}
\label{odd1}
{1\over r^3}\left( r^2\ f_5(r) \right)'= {2\over r^2}-{2\over 3}\left[ \Lambda -8\pi G\beta^2 \left(1-\sqrt{1+{3\eta^2\over \beta^2 r^2}} \right)\right],
\end{equation}
which can be integrated to yield
\begin{eqnarray}
\label{odd2}
f_5(r)&=&1-\frac{2 G M}{r^2}-\frac{\Lambda r^2}{6}+\frac{2}{3} \pi  G \bigg[-3 \eta ^2 \sqrt{\frac{3 \eta ^2}{r^2 \beta ^2}+1}-2 r^2 \beta ^2 \left(\sqrt{\frac{3 \eta ^2}{r^2\beta ^2}+1}-1\right)\nonumber\\
   &&+\frac{9 \eta ^4 \log \left(r \left(\sqrt{\frac{3 \eta ^2}{r^2 \beta ^2}+1}+1\right)\right)}{r^2 \beta ^2}\bigg].
\end{eqnarray}
The non-polynomial (logarithmic) nature of the solution makes it cannot be probed by means of~\eqref{eq:000}. We can better understand this solution by expanding it in $1/r$ but keeping $\beta$ small,
\begin{equation}
\label{odd3}
f_5(r)\approx1-4\pi G\eta^2-{2 G M\over r^2}-{\Lambda r^2\over6}-{3G\pi\eta^4\over2\beta^2 r^2}\left[1+12\log\left(2r\right)\right]+{9\pi G\eta^6\over2\beta^4 r^4}+\mathcal{O}\left(r^{-6}\right).
\end{equation}
It is manifest that the deficit solid angle is given by $4\pi G\eta^2$, following~\eqref{eq:delta}. The solution has singularity at the origin but regular everywhere else. To investigate the black hole horizons, we can follow condition~\eqref{eq:extremecondition}. The mass is given by
\begin{eqnarray}
M_5&=&\frac{1}{12 G\beta ^2}\bigg[8\pi  G r_+^4 \beta ^4-12 \pi  G r_+^2 \beta ^2 \eta ^2
   \sqrt{\frac{3 \eta ^2}{r^2 \beta ^2}+1}+36 \pi  G \eta ^4 \log
   \left(r_+ \left(\sqrt{\frac{3 \eta ^2}{r_+^2 \beta
   ^2}+1}+1\right)\right)\nonumber\\
   &&-8 \pi  G r^4 \beta ^4 \sqrt{\frac{3 \eta
   ^2}{r_+^2 \beta ^2}+1}-r_+^4 \beta ^2 \Lambda +6 r_+^2 \beta ^2\bigg],  
\end{eqnarray}
while the extremal radius $r_e$ satisfies
\begin{equation}
8 \pi  G r_e^2 \beta ^2 \sqrt{\frac{3 \eta ^2}{r_e^2 \beta ^2}+1}-8 \pi 
   G r_e^2 \beta ^2+r_e^2 \Lambda -3=0.
\end{equation}   
For non-zero $\Lambda$ there are at most two possible horizons, and they merge at $r=r_e$ in the extremal condition.

This analysis can be generalized for $D=7, 9$, or any odd dimensions.

   
\section{Thermodynamics of black holes with global defects: A review}\label{sec:thermoreview}

Black hole is a thermodynamical object; {\it i.e.,} it radiates and undergoes phase transition. Black hole with global defects is no exception. We investigate the thermodynamical properties of the black hole solutions obtained above. The results are genuine. The nonlinearity of DBI structure considerably modifies the thermodynamical stability of the black hole. In order to fully appreciate such deviation, in this section we shall review the thermodynamical properties of Barriola-Vilenkin black hole.

The thermodynamics of $4d$ black holes with global monopole has been studied with various approach since early 1990s \cite{Harari:1990cz,Jing:1993np,Yu:1994fy,Jensen:1994yz,Zhao:2005ej}. The proposal to consider the thermodynamics properties of black holes with global monopole was triggered when Harari and Lousto discussed the implication of global monopole's effective negative mass to the Hawking temperature of an evaporating black hole \cite{Harari:1990cz}. They argued that the $\Delta$ parameter would not affect the surface gravity of the black hole, hence leaving its contribution to be insignificant. Jing et al showed that the Bardeen-Carter-Hawking (BCH) law should be modified for the black holes with global monopole, because the deficit angle makes the rate change of black hole's energy unequal to the rate change of its mass \cite{Jing:1993np}. They discovered that the entropy of the black hole is still consistent with Bekeinstein-Hawking's result, $S=\frac{1}{4}A$. Later, Yu laid the foundation for thermodynamics calculation of black holes with global monopole by employing Euclidean action and surface gravity method \cite{Yu:1994fy}. He found that a configuration of black holes with global monopole is thermodynamically feasible, as long as the mass of the Schwarzschild black holes is larger than the mass of global monopole. He also found that the Hawking temperature of the black hole should be lowered because of the deficit angle, thus leaving its horizon area larger. This would not violate the $2^{nd}$ law of black hole mechanics, because the entropy-area relation remains the same. Another interesting finding from the thermodynamics of black holes with global monopole is the fact that its entropy could be expressed as Cardy-Verlinde formula \cite{Zhao:2005ej}. Based on works that have been conducted in this framework, we briefly review the thermodynamics of higher dimensional black holes with global defects.

\begin{figure}
\begin{tabular}{cc}
  \includegraphics[width=8cm]{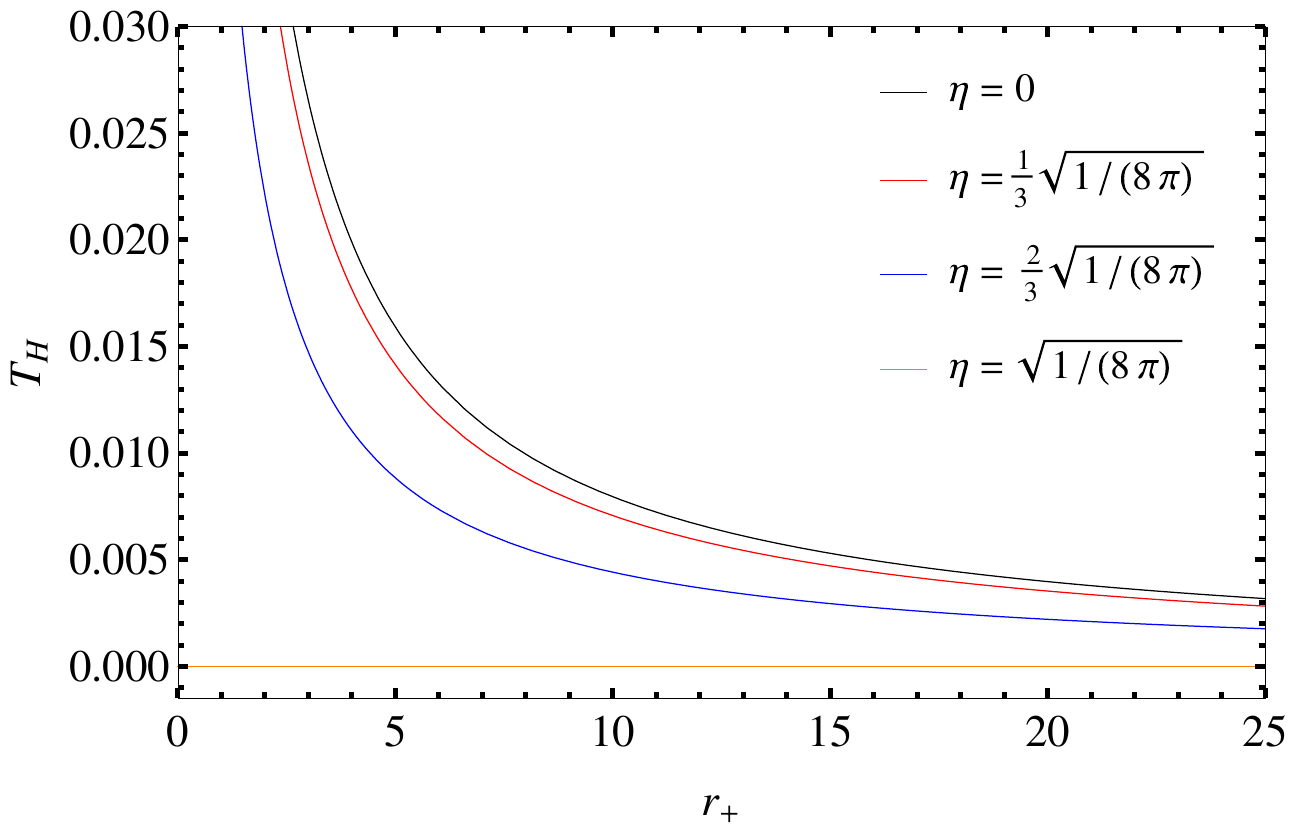} &   \includegraphics[width=8cm]{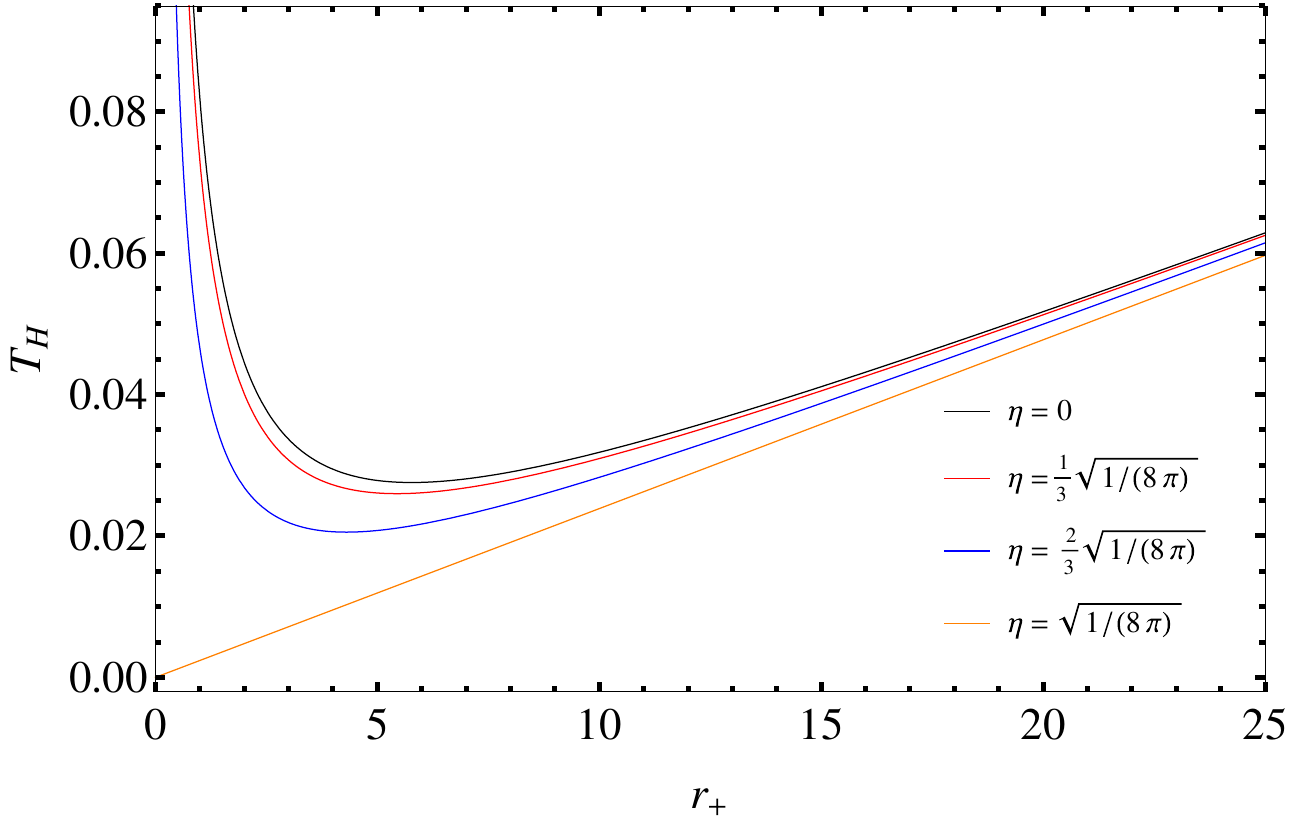} \\
(a) & (b)
\end{tabular}
\caption{A typical temperature as a function of event horizon in $D=4$: $(a) \Lambda=0$,  $(b) \Lambda=-0.03$.}
\label{fig:00nTvr}
\end{figure}

The temperature of black holes can be obtained from the relation $T_{H} = \kappa/2\pi$, also known as the Hawking temperature \cite{Hawking:1974sw}. The surface gravity for spherically symmetric black holes is given by \cite{Visser:1992qh}
\begin{equation}
\kappa = \left. \frac{1}{2}\frac{\partial_{r}g_{00}}{\sqrt{-g_{00}g_{rr}}} \right\vert_{r_{h}}.\label{eq:0k0}
\end{equation}
We use the modified Tangherlini metric to obtain the surface gravity (\ref{eq:expandbars}). From the Hawking relation, the temperature is given by
\begin{equation}
T_{H} = \frac{D^2 - D \left(8 \pi  \eta ^2 G+5\right)+2 \left(8 \pi  G \eta^{2} +\Lambda  r_{+}^2-3\right)}{4 \pi r_{+} (D-2)}.\label{eq:0TH0}
\end{equation}
We show the effect of global defects hair on the Hawking temperature of a black hole in Fig. \ref{fig:00nTvr}. We can see that the essential thermodynamic characteristics of Tangherlini black holes still appear in this global defects case, both in asymptotically-flat and AdS (has a local minimum $T_{0}$ at $r_{+} = r_{0}$) spacetime.

Here we compare the stability of several black holes as its temperature reaches the Hawking limit with different value of deficit angle. The Hawking temperature is lowered as the value of deficit angle increases, and if a black hole started out with maximum amount of deficit angle then it is in the extremal state in the first place ($T_{H} = 0$). Extremal state is the most stable configuration a black hole could have with minimum amount of cosmological "hair". Hence, black holes with the least amount of global monopole would be more unstable than black holes with maximum amount of global monopole.

The entropy of black holes can be obtained by using thermodynamics relation 
\begin{equation}
S = \int dr_{+} \left(1/T_{H}\right)\partial_{r_+} M,
\end{equation}
which comes from the first law of black hole thermodynamics. Entropy of black holes with global monopole is found to be
\begin{equation}
S = \frac{2 \pi  r_{+}^{D-2}}{G (D-2)}.\label{eq:0S0}
\end{equation}
This is consistent with the entropy of Schwarzschild black hole, i.e. $S_\text{Sch} = \pi r_{+}^2/G$ (for $D=4$), which is proportional to the radius, of the horizon, squared.

Specific heat is one the important quantities in black hole mechanics to measure the local stability of the black hole configuration. The sign of specific heat indicates the stability of black hole in certain phases (see, for example, Ref~\cite{Roychowdhury:2014cva} and the references therein). The singularity in the specific heat signals the phase transition. The specific heat of a black hole can be obtained through the relation $C_H= T_H \left(\partial S/\partial T_H \right)$, and the result for our black holes is as follows
\begin{equation}
C_H= \frac{- 2 \pi r_{+}^{D-2}\left(D^2 - D\left(5 + 8 \pi G\eta^{2} \right) + 16 \pi G \eta^{2}  - 2\Lambda r_{+}^{2} + 6 \right)}{G \left(D^2 - D\left(5 + 8 \pi G\eta^{2} \right) + 16 \pi G \eta^{2}  + 2\Lambda r_{+}^{2} + 6 \right) }.\label{eq:0C0}
\end{equation}
\begin{figure}
\begin{tabular}{cc}
  \includegraphics[width=8cm]{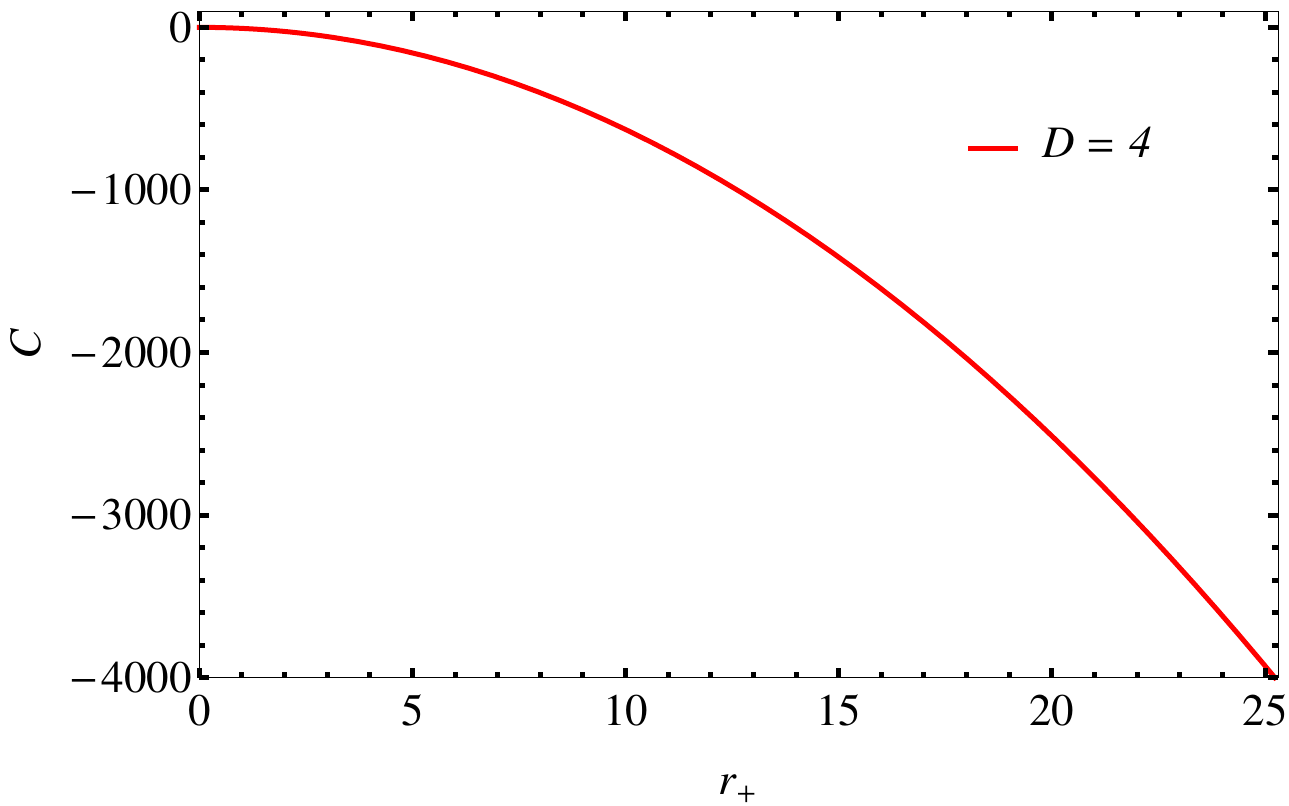} &   \includegraphics[width=8cm]{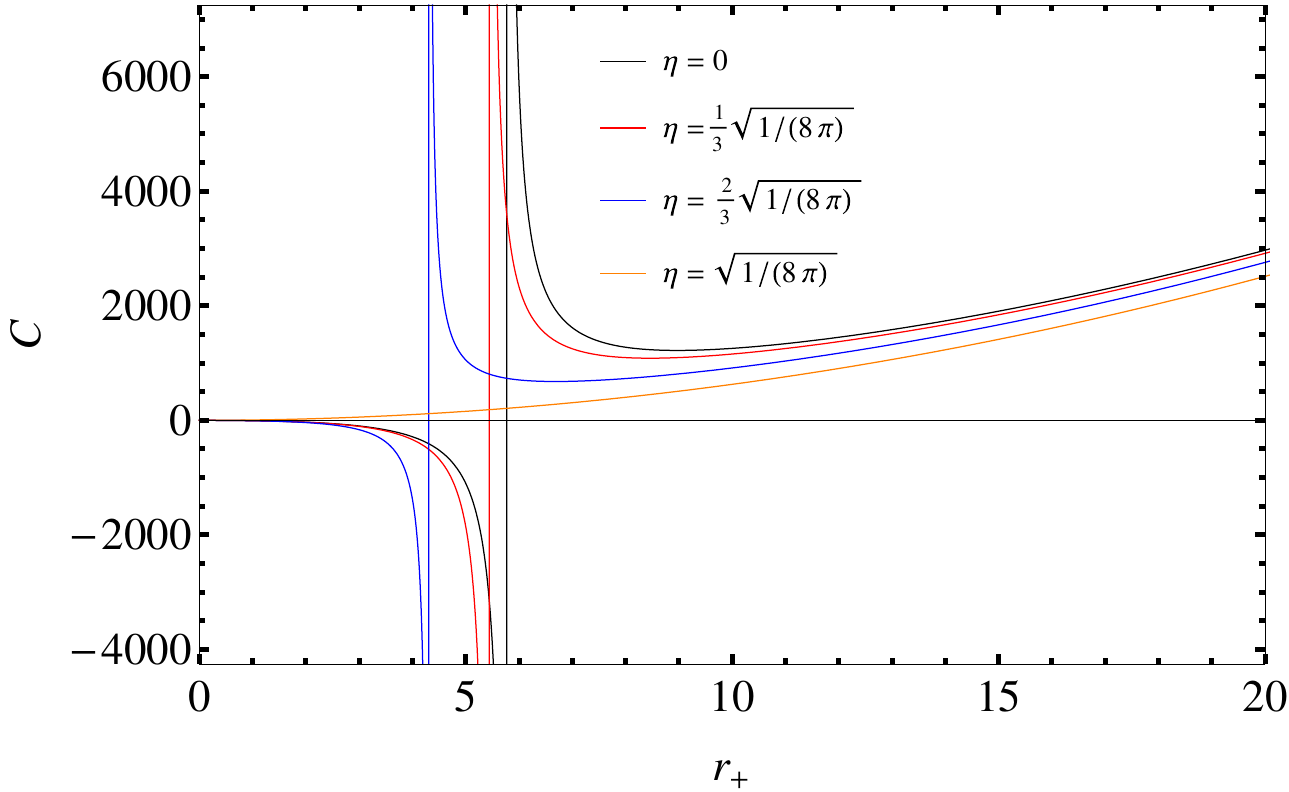} \\
(a) & (b) \\[6pt]
\includegraphics[width=8cm]{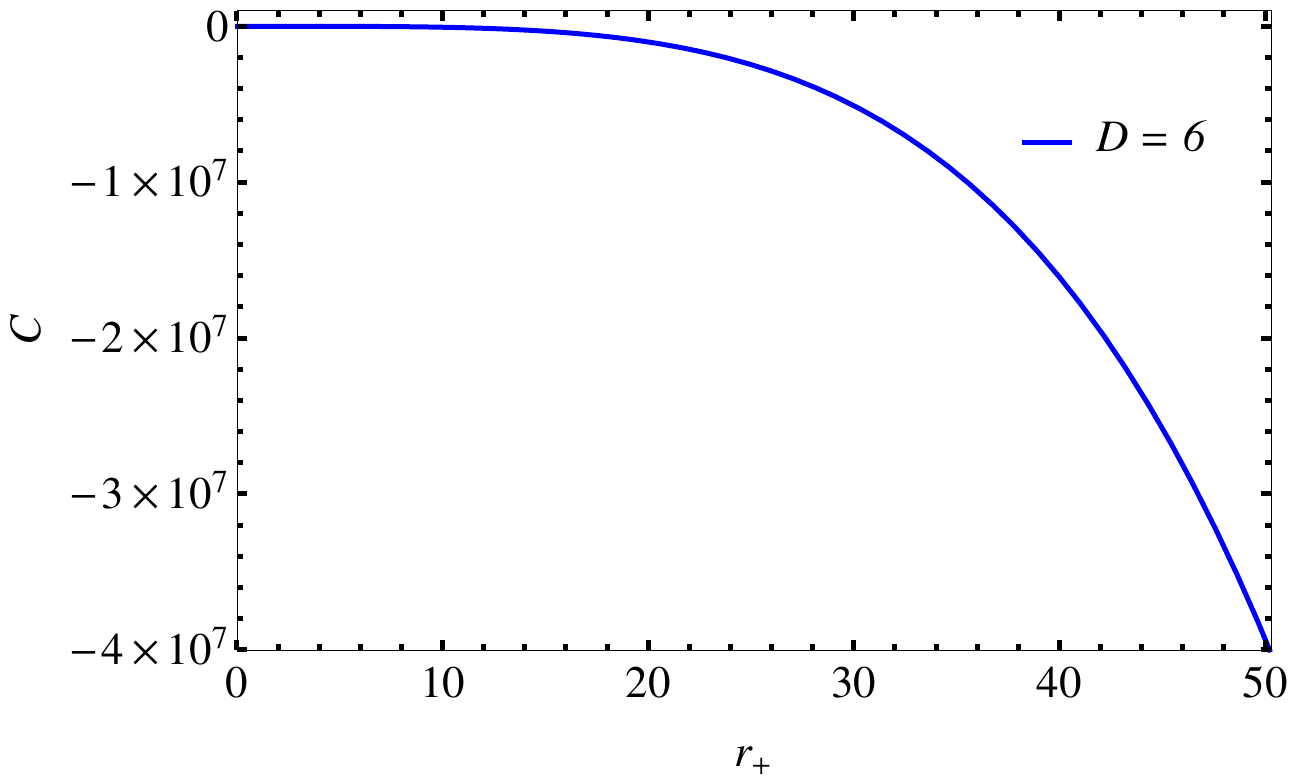} &   \includegraphics[width=8cm]{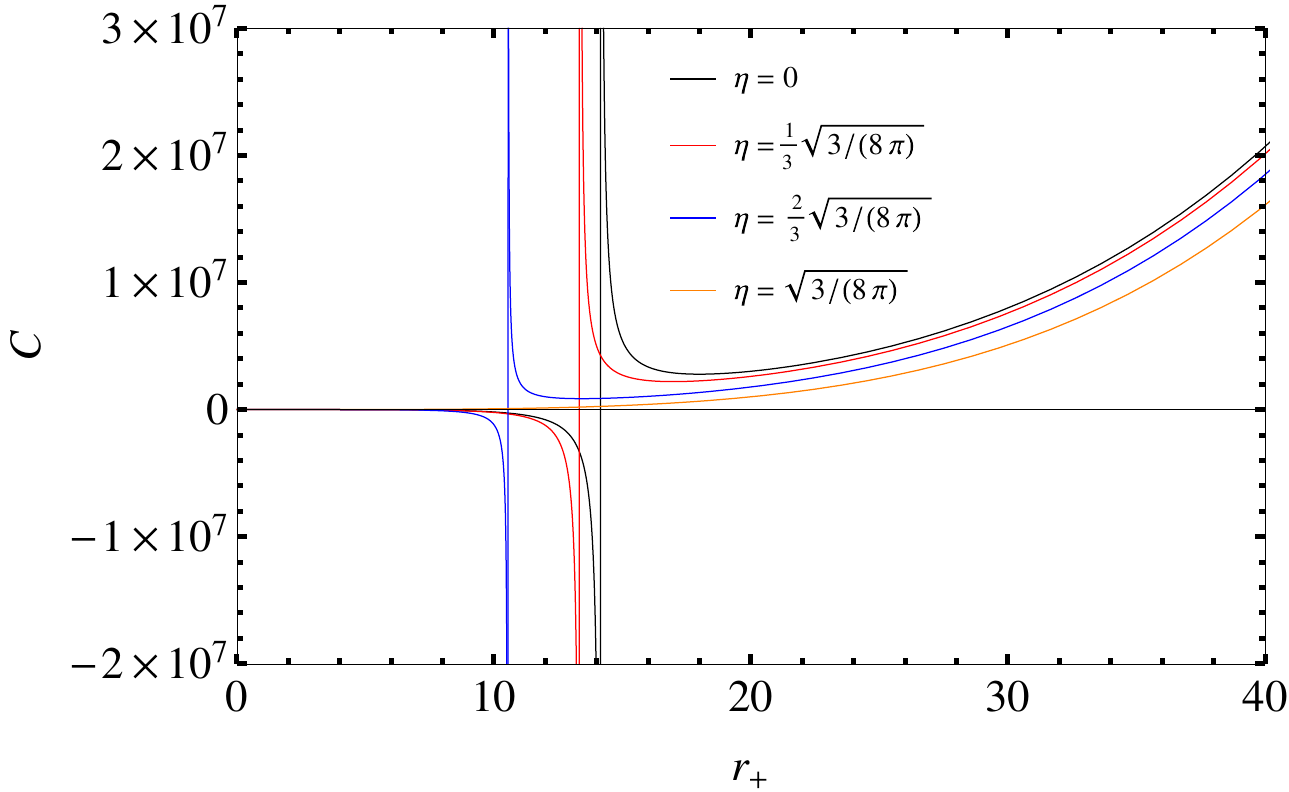} \\
(c) & (d) \\[6pt]
\end{tabular}
\caption{A typical specific heat as a function of event horizon, with $(a)~D=4$ and $\Lambda=0$, $(c)~D=6$ and $\Lambda=0$, $(b)~D=4$ and $\Lambda=-0.03$, $(d)~D=6$ and $\Lambda=-0.03$.}
\label{fig:00nCvr}
\end{figure}
Typical profiles for this function is shown in Fig.~\ref{fig:00nCvr}. When the calculation is conducted in asymptotically-flat spacetime, the specific heat would not depend on the value of deficit angle because the formula would be reduced into the typical asymptotically-flat Tangherlini form ($C_H = - 2 \pi r^{D - 2}/G$). Because of that, the effect from different values of deficit angle in flat spacetime would be indistinguishable once we analyze the specific heat.

\section{Thermodynamics of black holes with DBI global defects}\label{sec:thermodbi}

Having reviewed the thermodynamical properties of black hole with ordinary global defects, in this section we wish to compare such results with that of black holes with DBI global defects. We will conduct the calculation with surface gravity method \cite{Yu:1994fy}. After obtaining the needed thermodynamics quantities, we can analyze the stability and phase transition of the black holes. 

Inserting the mass equation \eqref{eq:massex} into the metric solution \eqref{eq:004}, which is later substituted into \eqref{eq:0k0}, the surface gravity can straightforwardly be obtained which then gives the temperature for our model as follows
\begin{equation}
T_{H} = \frac{D^2+ D \left(8 \pi  \eta ^2 G-5\right)-2 \left(8 \pi  G \left(\eta
   ^2+\beta ^2 r_{+}^2 \left(\sqrt{\frac{\eta ^2 (D-2)+\beta ^2 r_{+}^2}{\beta ^2
   r_{+}^2}}-1\right)\right)+\Lambda  r_{+}^2-3\right)}{4 \pi r_{+} (D-2)}.\label{eq:TH0}
\end{equation}

The entropy of black holes with DBI global defects is found to be identical with \eqref{eq:0S0}. Thus, the entropy is independent on both $\eta$ and $\beta$, as well as on the dimensionality. This is not surprising and is consistent with the previous result\footnote{It is quite interesting, though, that the entropy of $4d$ black hole with DBI scalar hair is independent on $\beta$, since its massive DBI global monopole counterpart is said to have bigger absolute value of effective negative mass~\cite{Liu:2009eh}.}~\cite{Jing:1993np, Yu:1994fy}. 

From $C_H = T_H \left(\partial S/\partial T_H \right)$, the specific heat expression of our model is as follows
\begin{equation}
C_H = \frac{2 \pi r_{+}^{D-2}\left( D^2 - D\left(5 - 8 \pi \eta^{2} G\right) - 16 \pi G \left(\eta^{2} + \beta^{2}r_{+}^{2}\left( \sqrt{\frac{\eta ^2 (D-2)+\beta ^2
   r_{+}^2}{\beta ^2 r_{+}^2}}-1 \right)\right) + 2\Lambda r_{+}^{2} - 6 \right)}{8 \pi  G^2 \left(\eta ^2 (2-D)+2 \beta ^2 r_{+}^2 \left(1-\frac{1}{\sqrt{\frac{\eta
   ^2 (D-2)+\beta ^2 r_{+}^2}{\beta ^2 r_{+}^2}}}\right)\right)-G \left(-D^2+5 D+2 \Lambda  r_{+}^2-6\right)}.\label{eq:C0}
\end{equation}

Based on the constraint found in the second section, we shall do the thermodynamics analysis for black holes with even-numbered dimension, specifically $D > 3$, up until $D=8$, and also $D=5$ as the odd-numbered dimension case. Additionally, we will compare the thermodynamic properties of these black holes with various value of $\beta$. Note that for simplicity in all of the subsequent calculations we have set $G=c=\hbar=1$.

\subsection{Asymptotically-flat black holes}

From equation (\ref{eq:TH0}), the temperature of this black holes with DBI global monopole ($D=4$) is given by
\begin{equation}
T_H = \frac{1 + 8 G \pi \left(\eta^2 - \beta^2 r_{+}^2\left(\sqrt{1 + \frac{2 \eta^2}{\beta^2 r_{+}^2}} - 1 \right)\right)}{4 \pi r_{+}}.\label{eq:TH1}
\end{equation}

\begin{figure}
  \includegraphics[width=8cm]{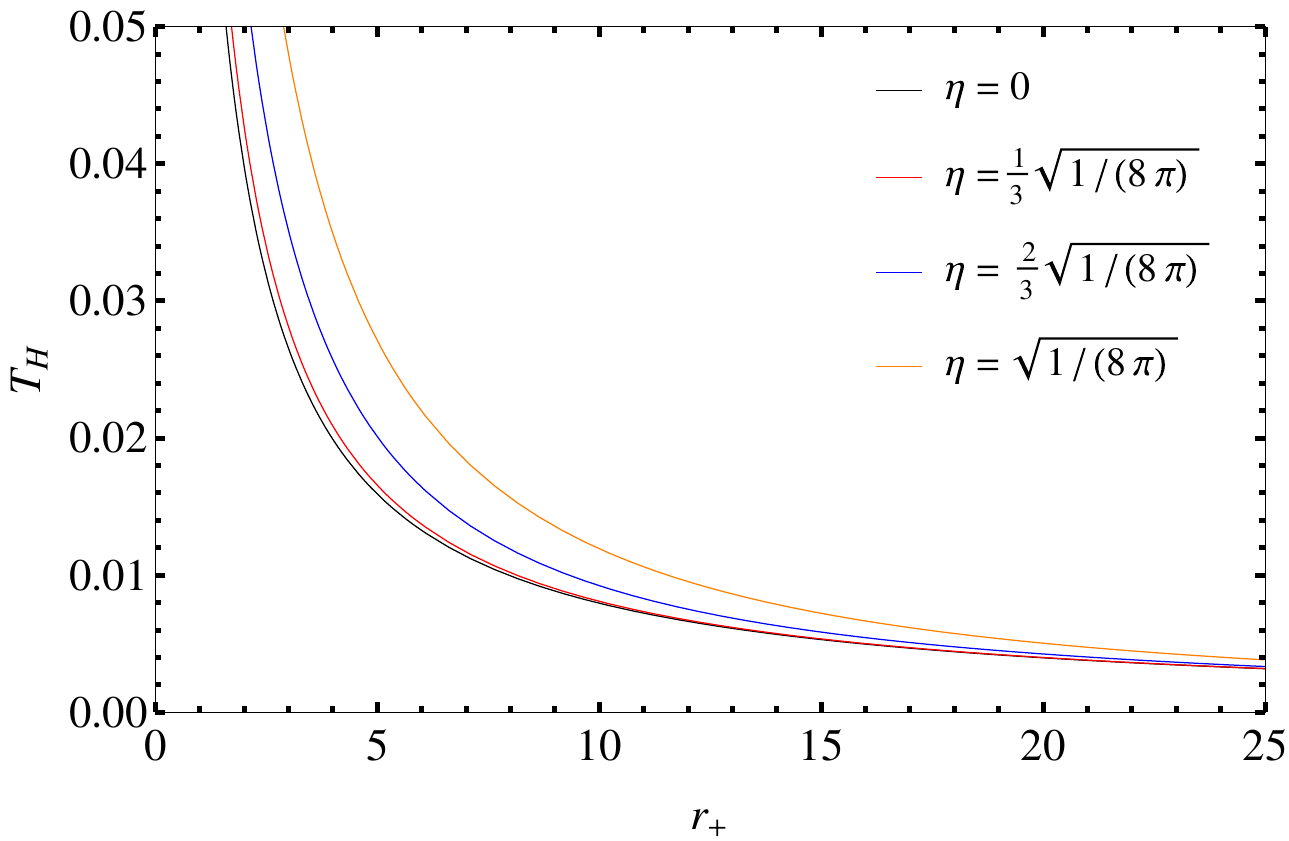}    \includegraphics[width=8cm]{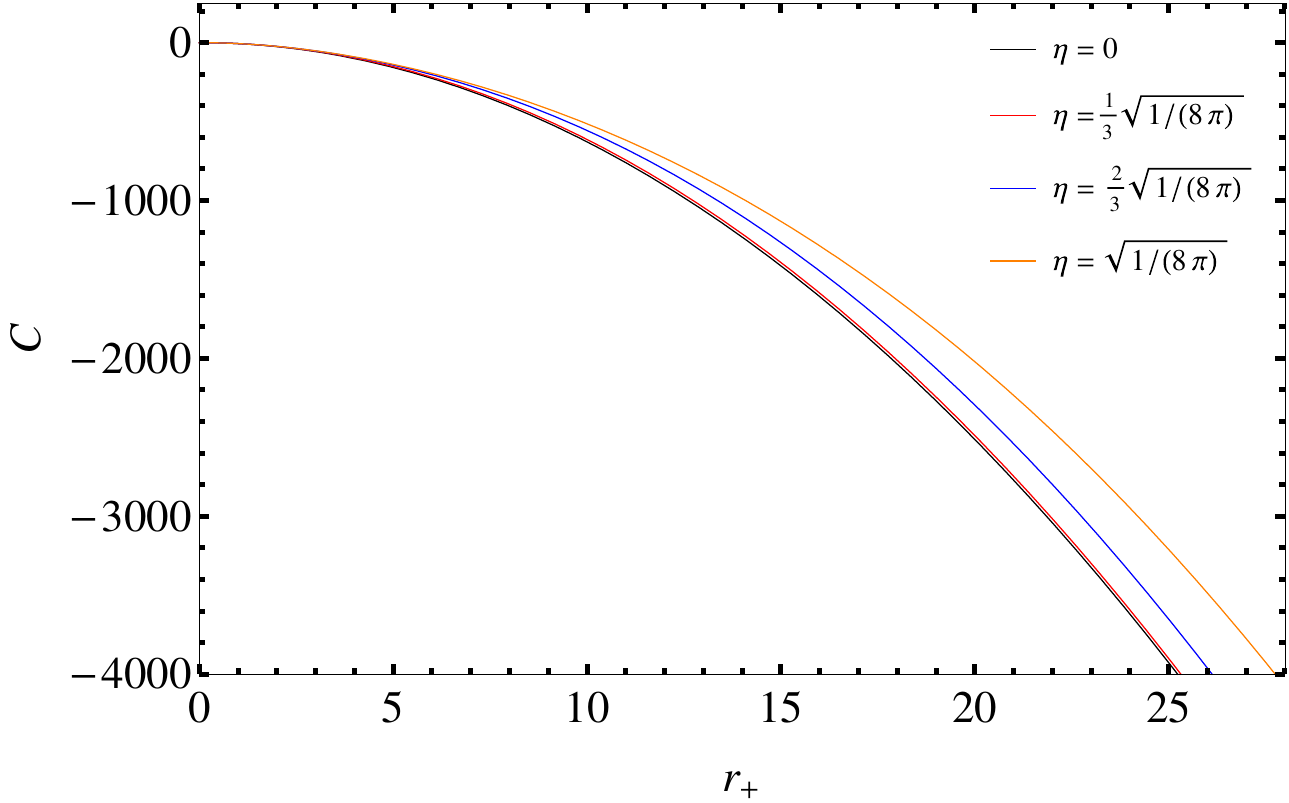}
\caption{Hawking temperature and specific heat as a function of event horizon for $\Lambda=0$ and $\beta=0.01$ in $D=4$.}
\label{fig:fn4Tvrfn4Cvr}
\end{figure}

We can see in Fig. \ref{fig:fn4Tvrfn4Cvr} that black holes with larger amount of global charge has higher temperature than the smaller ones. This is the case for $D=6$ and $D=8$ as well. These are in contrast to the case of black hole with (canonical) global defects discussed in the previous section. The DBI nonlinear structure has modified the thermodynamical configuration of BV black hole; the greater the scalar charge the more it radiates.


After obtaining the entropy for $D=4$ from (\ref{eq:0S0}), we can find the specific heat of the black holes, which is given by
\begin{equation}
C_H = \frac{2 \pi  r_{+}^2 \sqrt{\frac{2 \eta ^2}{\beta ^2 r_{+}^2}+1} \left(1 -8 \pi  \left(\beta ^2 r_{+}^2 \left(\sqrt{\frac{2 \eta ^2}{\beta ^2 r_{+}^2}+1}-1\right)-\eta
   ^2\right)\right)}{8 \pi  \left(\beta ^2 r_{+}^2 \left(\sqrt{\frac{2 \eta ^2}{\beta ^2 r_{+}^2}+1}-1\right)-\eta ^2 \sqrt{\frac{2 \eta ^2}{\beta
   ^2 r_{+}^2}+1}\right)-\sqrt{\frac{2 \eta ^2}{\beta ^2 r_{+}^2}+1}}.\label{eq:C11}
\end{equation}
We plot the specific heat of black holes in $D=4$. As can be seen in the Fig. \ref{fig:fn4Tvrfn4Cvr}, the specific heat will always be negative. The same qualitative property appears as well in $D=6, 8$. This behavior is similar to the usual Schwarzschild (or Tangherlini), which is known to be locally thermodynamically unstable \cite{Altamirano:2014tva}. This unstable phase can be related to the instability of flat spacetime itself, that has been studied extensively in \cite{Myung:2007my, Gross:1982cv}.



\begin{figure}
  \includegraphics[width=8cm]{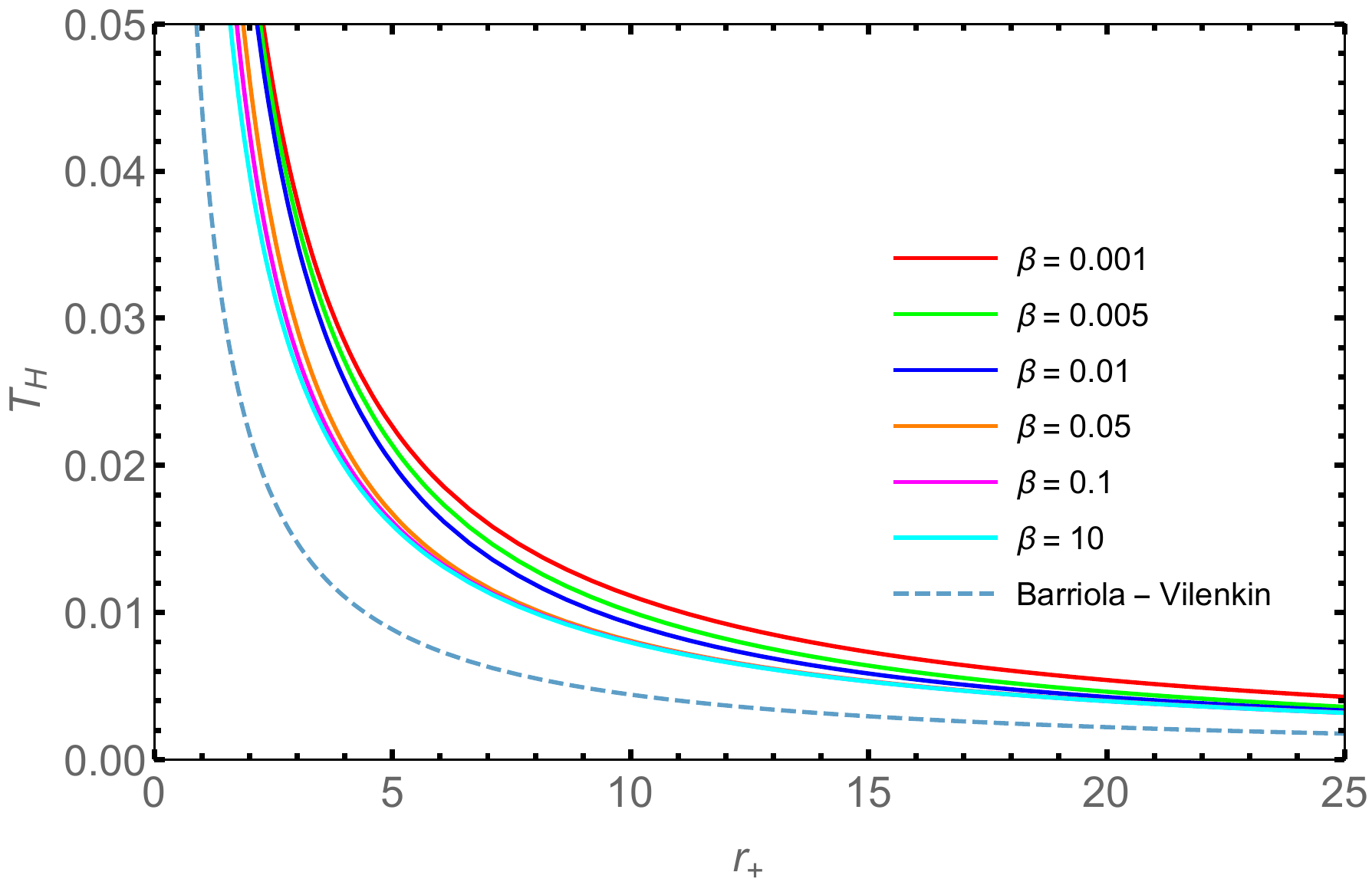}    \includegraphics[width=8cm]{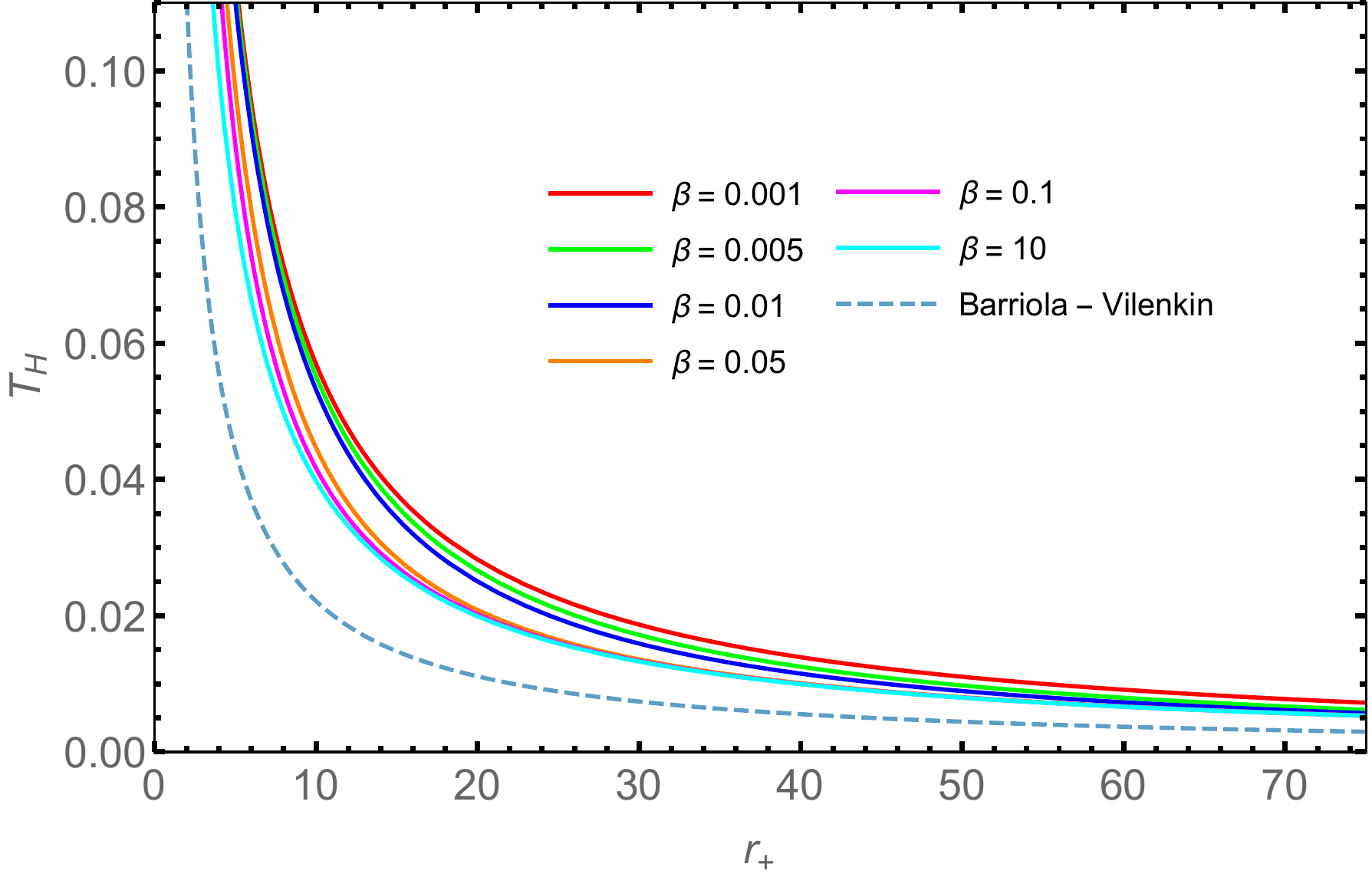}
\caption{Hawking temperature as a function of event horizon for $\Lambda=0$, with $D=4, \beta=0.133$(left) and $D=8, \beta=0.297$ (right).}
\label{fig:betafnTvr}
\end{figure}

\begin{figure}
  \includegraphics[width=8cm]{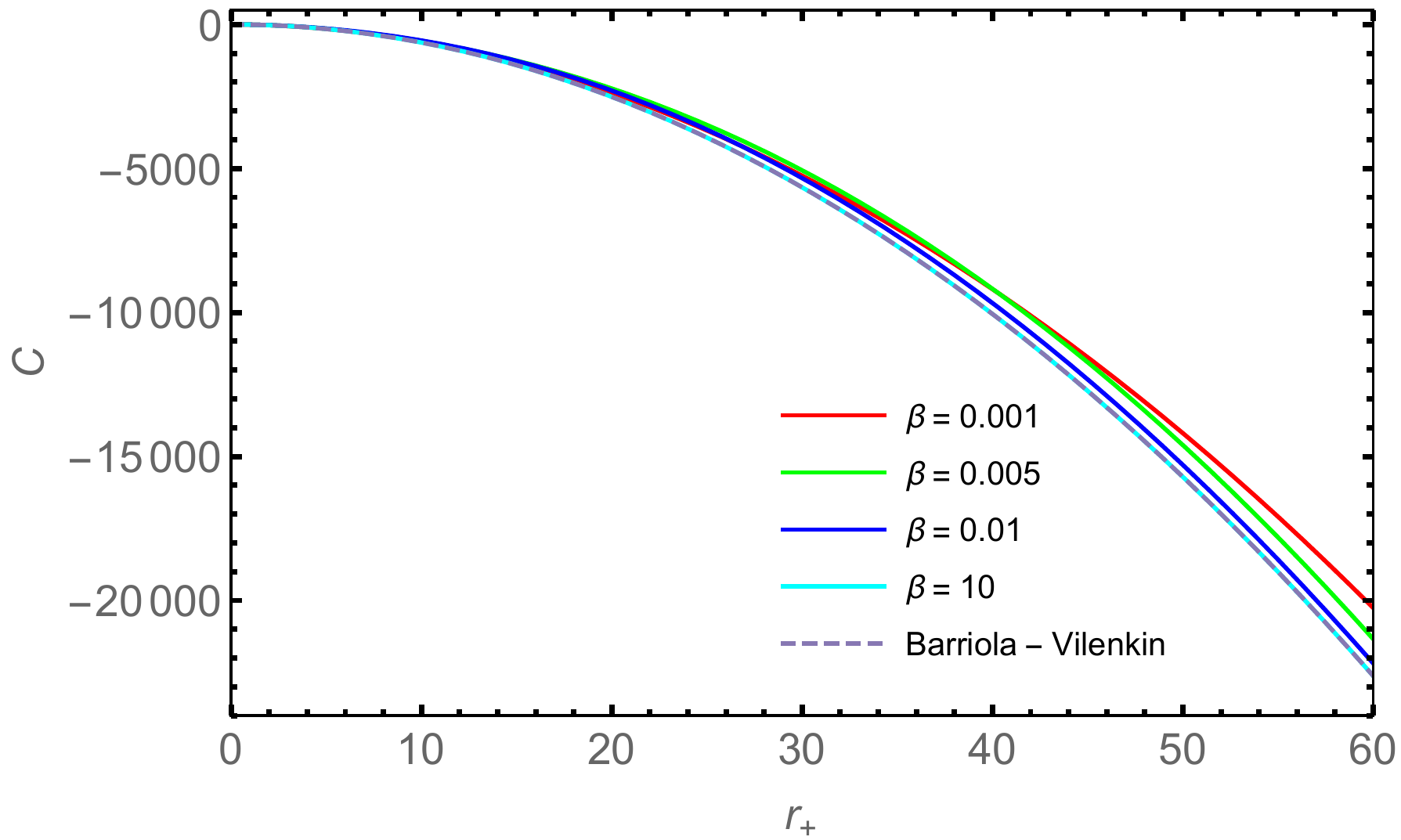}    \includegraphics[width=8cm]{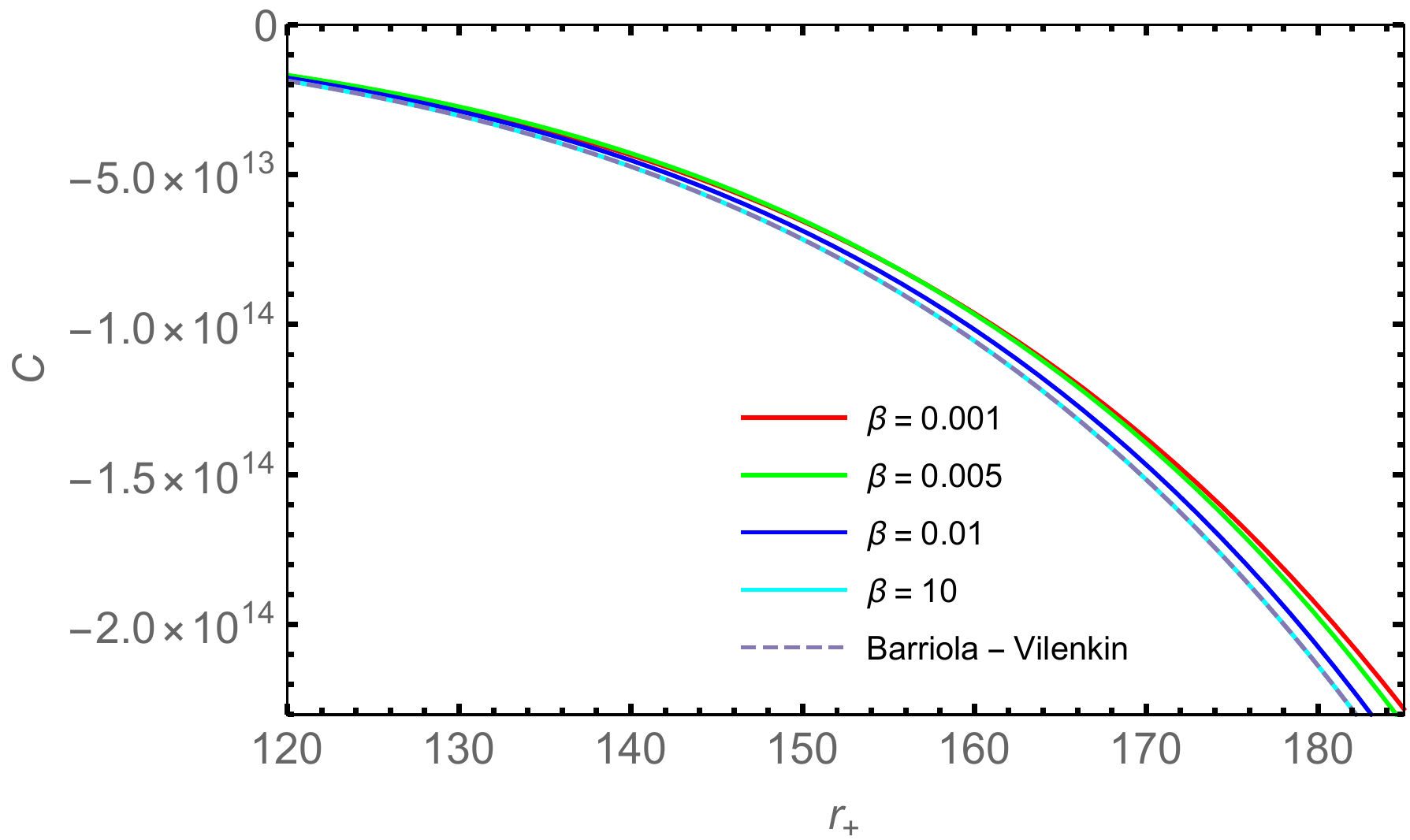}
\caption{Specific heat as a function of event horizon for $\Lambda=0$, with $D=4, \eta=0.133$ (left) and $D=8, \eta=0.297$ (right).}
\label{fig:betafnCvr}
\end{figure}

Based on the constraint found in Eq. \ref{chi}, the black hole with DBI global hair can be classified into strongly coupled ($\beta \ll 1$) and weakly-coupled ($\beta \gg 1$). We plot the temperature and  specific heat of black holes with various amount of $\beta$ in Fig. [\ref{fig:betafnTvr}, \ref{fig:betafnCvr}], respectively. It is evident that in every dimension, black holes with strongly coupled DBI global defects would have higher Hawking temperature than its weaker counterpart. Thus we can say that as the $\beta$ coupling becomes stronger, the black hole would radiate more. This results in higher specific heat in every dimension, although each configuration is still thermodynamically unstable from its negative-valued specific heat. 


In general, varying $\beta$ does not significantly alter the key feature of asymptotically-flat Tangherlini black holes with global monopole, in which the temperature increases as the black hole radius decreases and the only available phase is the unstable one. We can say that the effect of $\beta$ would be to shift the thermal stability of the black holes, in which stronger coupling results in higher temperature, and vice versa.

Almost in all cases, black holes with weakly coupled DBI global monopole have the closest value to BV black holes. This is to be expected, since our model should yield Barriola - Vilenkin result in the limit of $\beta \rightarrow \infty$. It is interesting to see that on the on-shell free energy diagram, as the radius increases, black holes with weakly coupled DBI global monopole and BV black holes becomes less adjacent, and eventually BV black holes will shift closer to the strongly coupled case.


\subsection{Anti de Sitter (AdS) black holes}

\subsubsection{$D=4$ case}

Different from (\ref{eq:TH1}), the Hawking temperature for non-flat black holes will have a term that includes the cosmological constant. The temperature of black hole with $D=4$ in this case is given by
\begin{equation}
T_H = \frac{1 + 8 G \pi \left(\eta^2 - \beta^2 r_{+}^2\left(\sqrt{1 + \frac{2 \eta^2}{\beta^2 r_{+}^2}} - 1 \right)\right) - r_{+}^2 \Lambda}{4 \pi r_{+}}.\label{eq:TH11}
\end{equation}


\begin{figure}[tp]
\begin{tabular}{cc}
  \includegraphics[width=8cm]{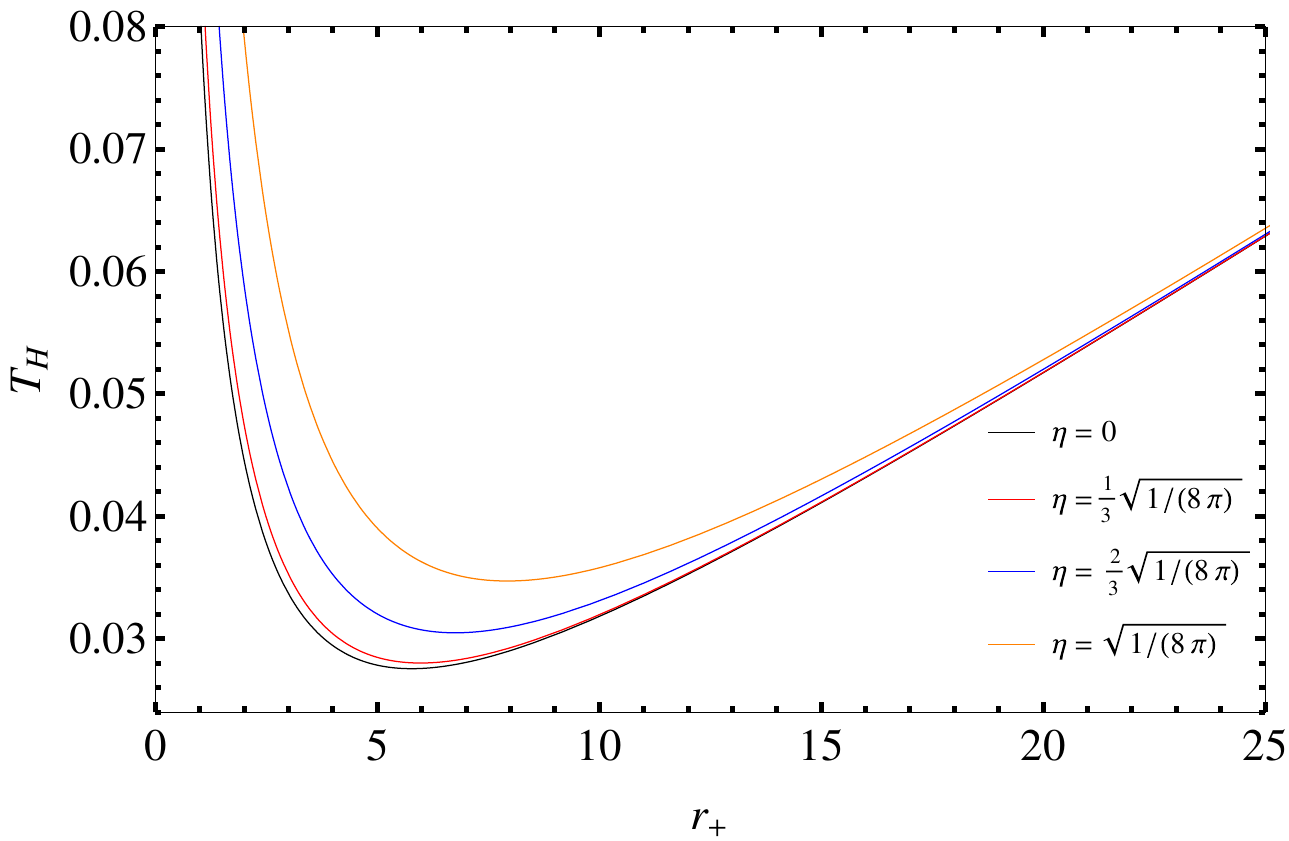} &   \includegraphics[width=8cm]{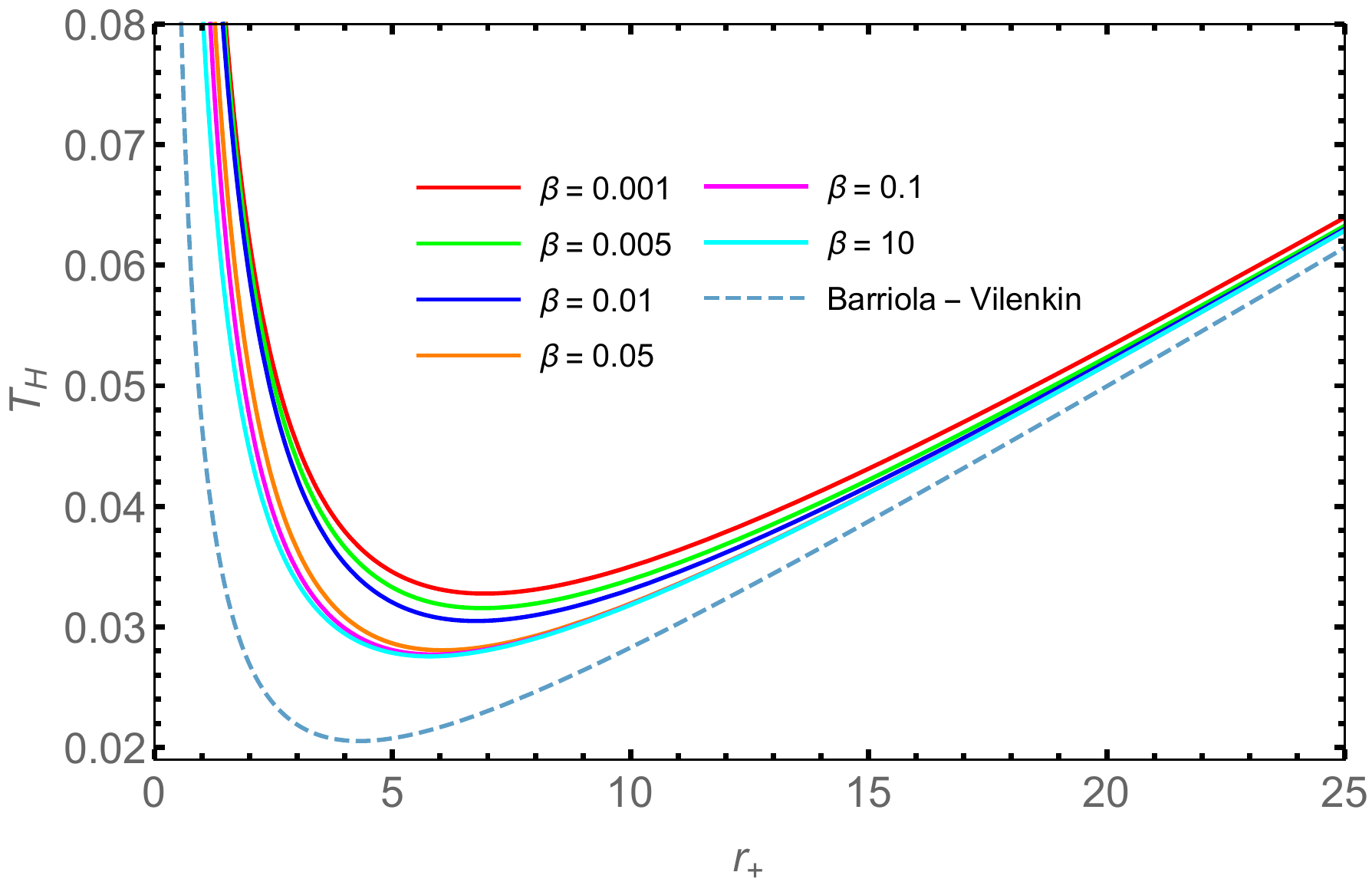} \\
(a) & (b) \\[6pt]
  \includegraphics[width=8cm]{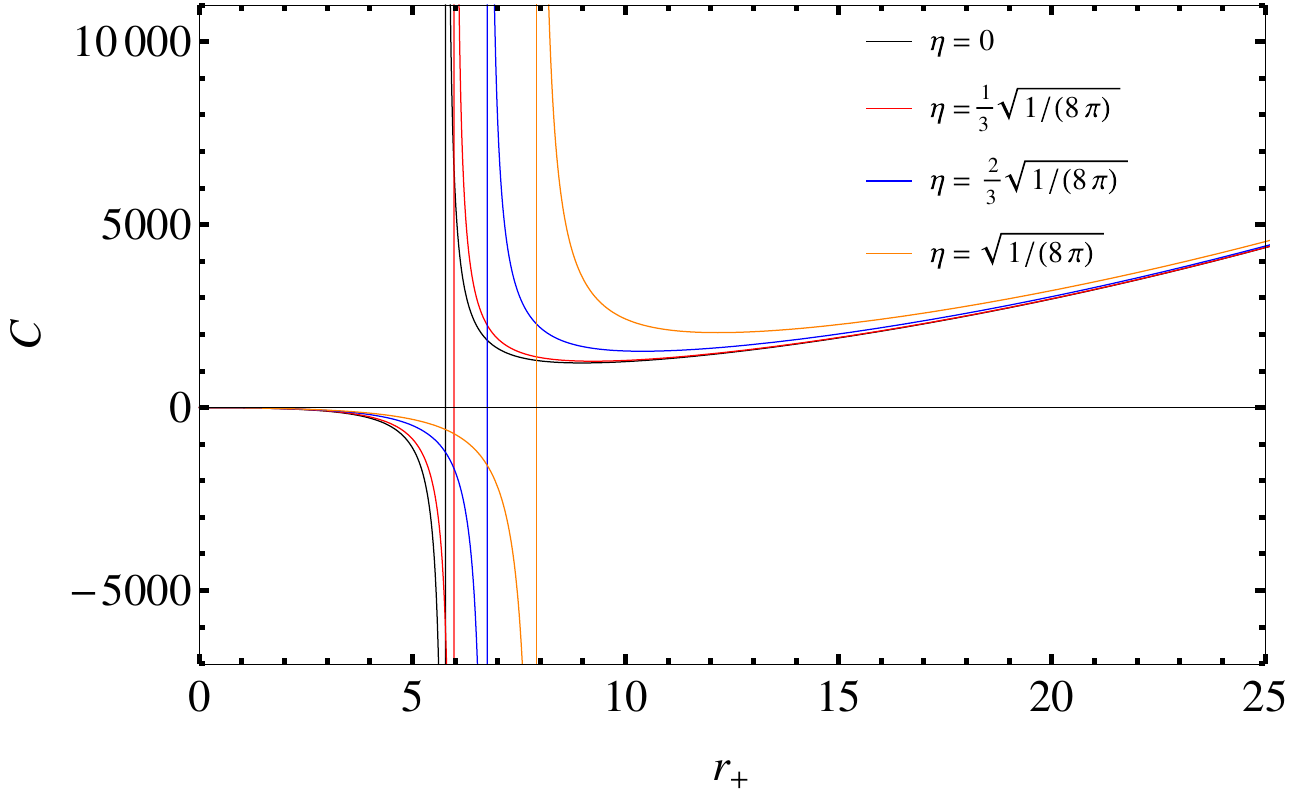} &   \includegraphics[width=8cm]{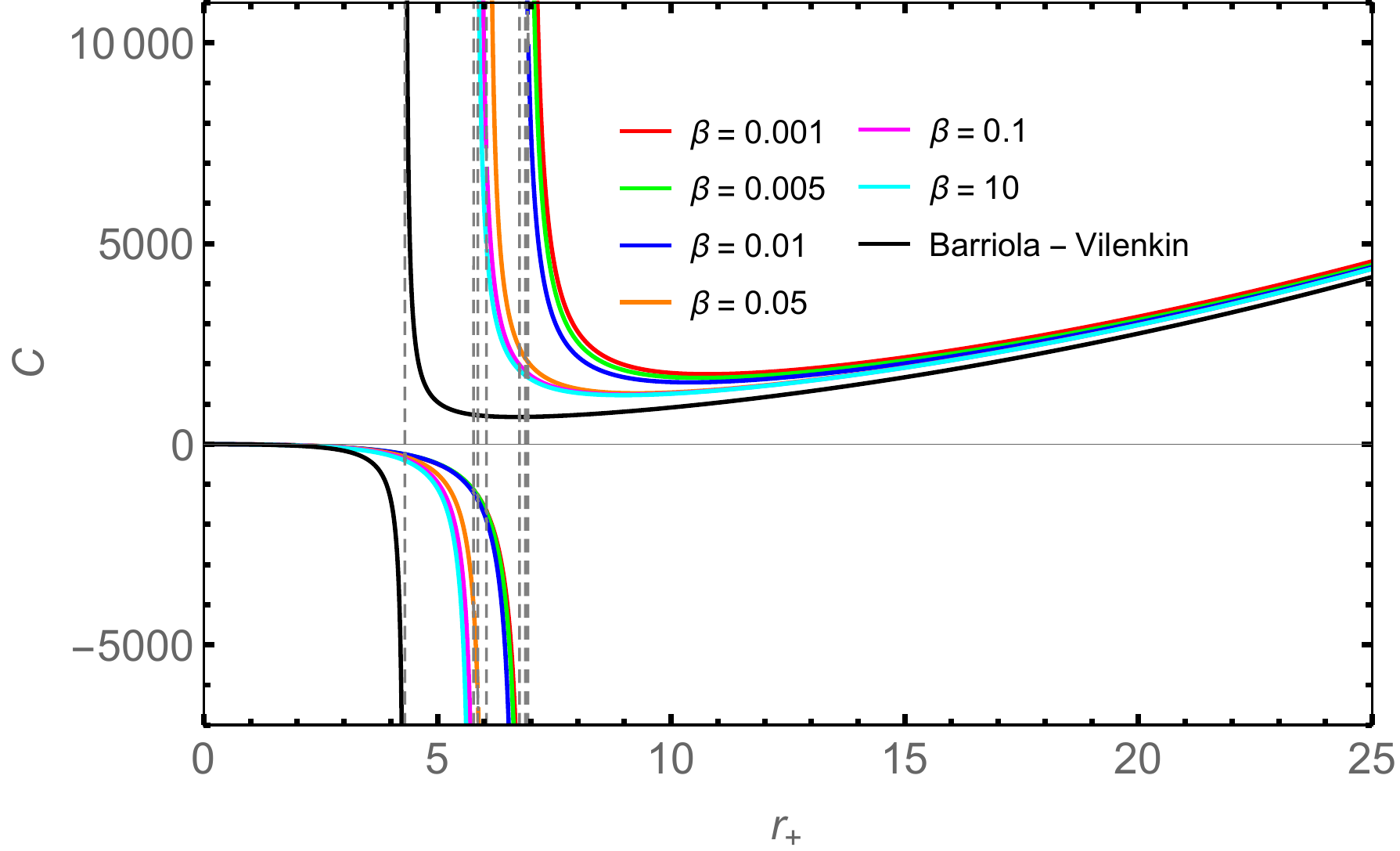} \\
(c) & (d) \\[6pt]
\end{tabular}
\caption{Hawking temperature as a function of event horizon, for $D=4$ AdS black holes with the value of $(a)\beta=0.01$ and $(b)\eta=0.133$, and also its specific heat as a function of event horizon, with $(c)\beta=0.01$ and $(d)\eta=0.133$.}
\label{fig:comparingn4Tvrn4Cvr}
\end{figure}

In the limit $\beta \rightarrow \infty $ and $\eta=0$, the temperature reduces to the usual AdS Schwarzschild black hole. Away from this limit, we plot in Fig. \ref{fig:comparingn4Tvrn4Cvr} the behavior of black hole temperature $T_H$ as a function of its horizon $r_+$ with various value of $\eta$ and $\beta$. The temperature behaves exactly like ordinary Schwarzschild AdS black hole, i.e. it has a local minimum at $r_{+} = r_{0}$ (where $\partial_{r_+}T_{H}|_{r_{+} = r_{0}}=0$). This local minimum will be shifted away when $\eta$ increases, and vice versa. This can be thought as the effect when the monopole alters the geometry of the black hole with its deficit angle, hence the surface gravity is modified by $\eta$. Varying $\beta$ results in higher temperature for the strongly-coupled case. Thus we can say that the non-linearity of $\beta$ renders the black hole to radiate more energy.



The entropy for $D=4$ black holes with DBI global monopole is the same as ordinary Schwarzschild black holes, $S = \pi r_{+}^{2}$. From equation (\ref{eq:C0}), the specific heat is given by
\begin{equation}
C_H = \frac{2 \pi  r_{+}^2 \sqrt{\frac{2 \eta ^2}{\beta ^2 r_{+}^2}+1} \left(1 -\Lambda  r_{+}^2 -8 \pi  \left(\beta ^2 r_{+}^2 \left(\sqrt{\frac{2 \eta ^2}{\beta ^2 r_{+}^2}+1}-1\right)-\eta
   ^2\right)\right)}{8 \pi  \left(\beta ^2 r_{+}^2 \left(\sqrt{\frac{2 \eta ^2}{\beta ^2 r_{+}^2}+1}-1\right)-\eta ^2 \sqrt{\frac{2 \eta ^2}{\beta
   ^2 r_{+}^2}+1}\right)-\left(\Lambda  r_{+}^2+1\right) \sqrt{\frac{2 \eta ^2}{\beta ^2 r_{+}^2}+1}}.\label{eq:C1}
\end{equation}
We show the behavior of specific heat as a function of event horizon with various value of $\eta$ and $\beta$ in Fig. \ref{fig:comparingn4Tvrn4Cvr}. The phase diagram structure of AdS black holes with DBI global monopole is also similar with AdS Schwarzschild black holes. There are two different phases for every Schwarzschild AdS black hole  configuration. Positive valued $C$ corresponds to a stable black hole state while the negative one corresponds to pure radiation state. The asymptotic vertical line corresponds to the critical point when the black holes undergo a first order phase transition, which is known as Hawking-Page transition \cite{Hawking:1982dh}. If the black hole temperature is less than a certain critical point, then the black hole will resolve into a pure thermal AdS state. The opposite happens when the black hole temperature is greater than the critical point, in which it will stays as a stable black hole.

The critical point in the variation of $\eta$ is shown as the radius when the black hole specific heat blows up to infinity, which is equal to $r_C = 5.77, 5.97, 6.75, 7.93$ as $\eta=0,{1\over3}\sqrt{1/8\pi},{2\over3}\sqrt{1/8\pi},\sqrt{1/8\pi}$, respectively, in the lower left panel of Fig. \ref{fig:comparingn4Tvrn4Cvr}. We can also see that from its stable phase, black holes with larger value of $\eta$ arrive at its critical point faster than the ones with smaller $\eta$, thus more prone to the instability than its smaller counterpart. If we consider the variation of $\beta$, it appears that increasing the strength of coupling to nonlinearity gives the same effect as increasing the symmetry-breaking scale $\eta$; it shifts the transition radius to the right. This means that for strongly-coupled DBI scalar hair, larger black holes can undergo phase transition.

\subsubsection{$D=6$ case}

\begin{figure}[tp]
\begin{tabular}{cc}
  \includegraphics[width=8cm]{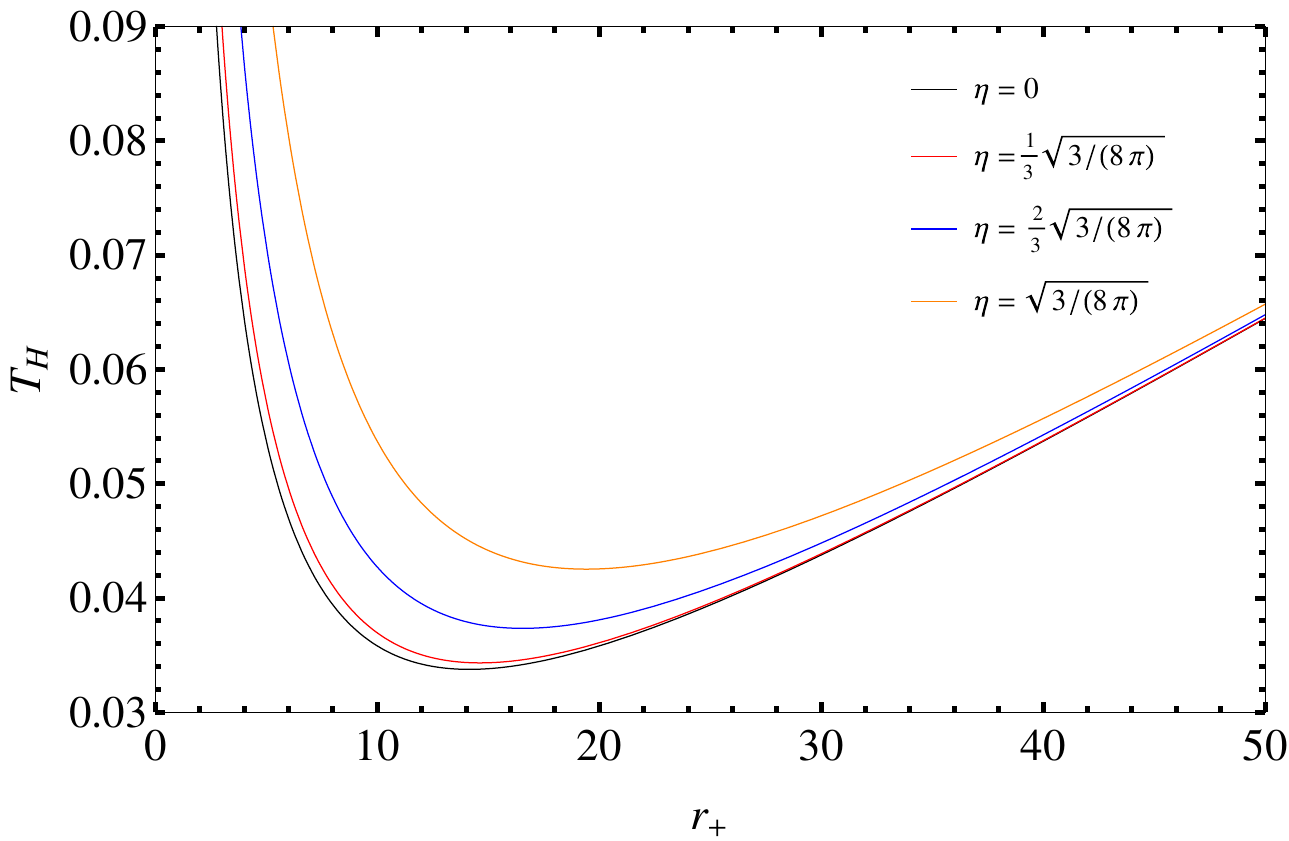} &   \includegraphics[width=8cm]{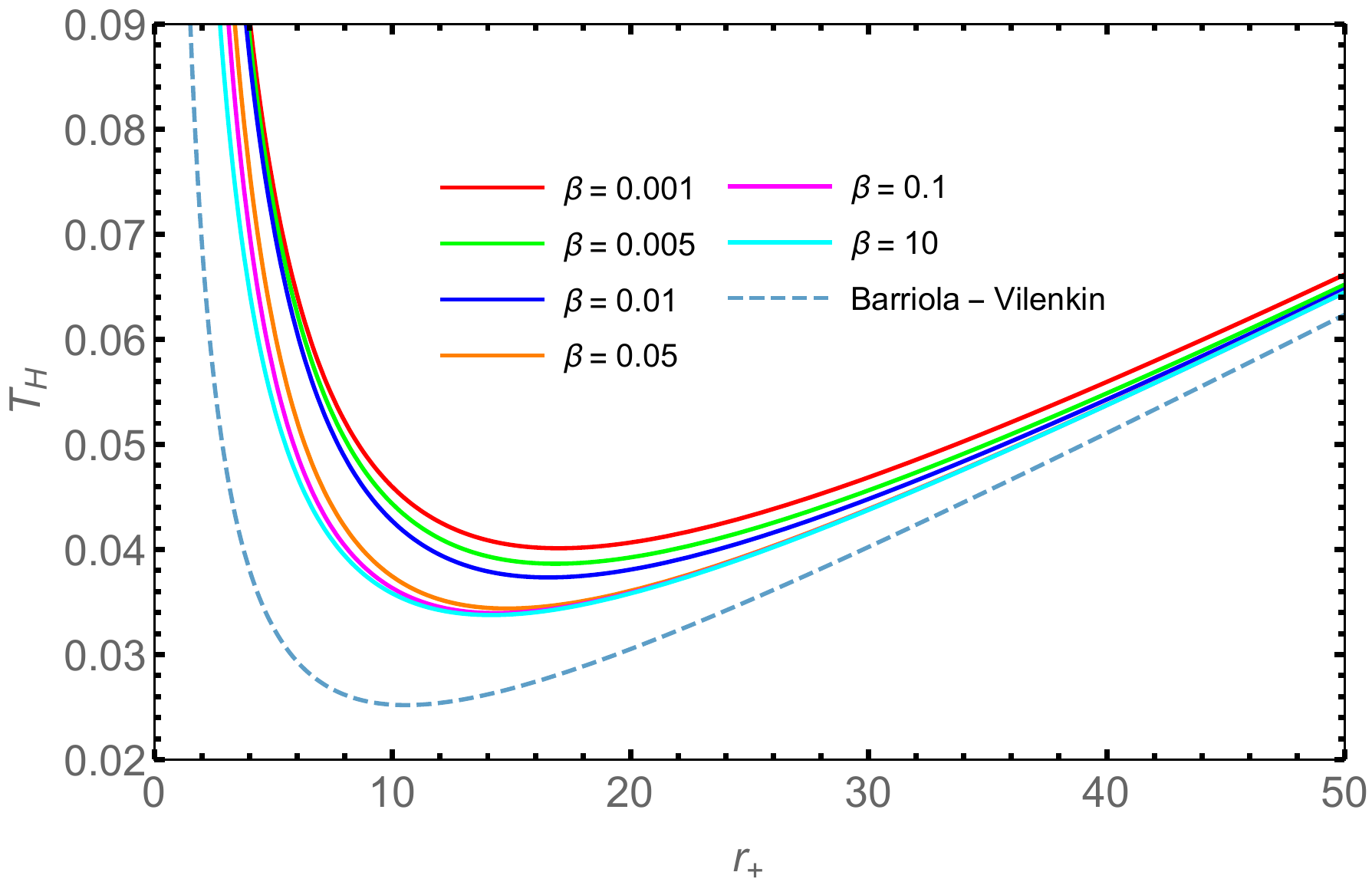} \\
(a) & (b) \\[6pt]
  \includegraphics[width=8cm]{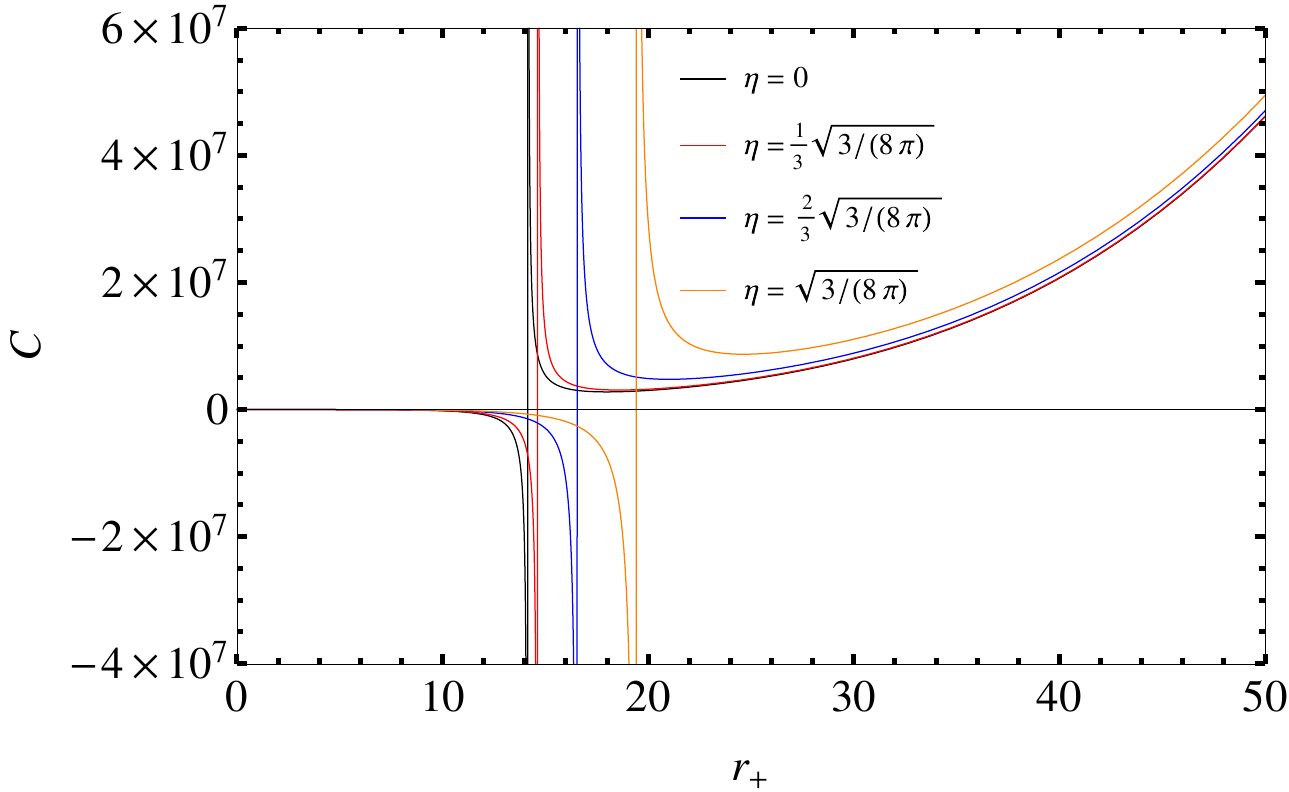} &   \includegraphics[width=8cm]{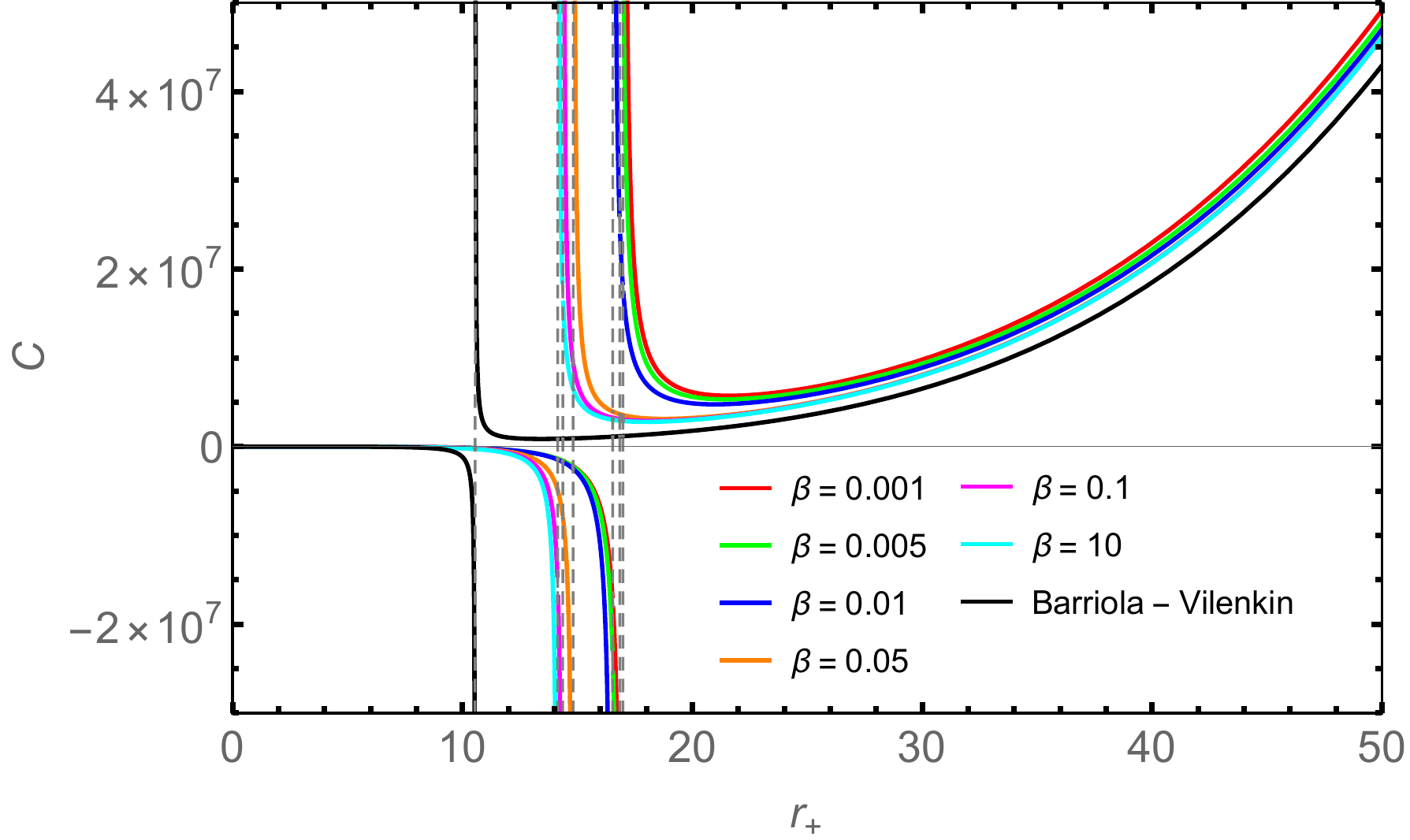} \\
(c) & (d) \\[6pt]
\end{tabular}
\caption{Hawking temperature as a function of event horizon, for $D=6$ AdS black holes with the value of $(a)\beta=0.01$ and $(b)\eta=0.230$, and also its specific heat as a function of event horizon, with $(c)\beta=0.01$ and $(d)\eta=0.230$.}
\label{fig:comparing6Tvrn6Cvr}
\end{figure}

From here our discussion on $D=6$ (and also $D=8$ later) will follow closely to the case of $D=4$.
From equation (\ref{eq:TH0}), the temperature of black hole with $D=6$ is given by
\begin{equation}
T_H = \frac{6 + \pi  \left(16 \eta ^2-8 \beta ^2 r_{+}^2 \left(\sqrt{\frac{4 \eta ^2}{\beta ^2 r_{+}^2}+1}-1\right)\right)-\Lambda  r_{+}^2}{8 \pi  r_{+}}.\label{eq:TH2}
\end{equation}
The behavior of black hole temperature with various value of $\eta$ and $\beta$ is shown in the lower panels of Fig. \ref{fig:comparing6Tvrn6Cvr}. Note that when $D=6$, the critical value of $\eta$ must be accordingly modified to a new value depending on $D$, which is $\eta_{crit}=\sqrt{3/(8\pi)}$.

The entropy for our $D=6$ black hole is $S = \pi r_{+}^{4}/2$. From equation (\ref{eq:C0}), the specific heat is given by
\begin{equation}
C_H = \frac{2 \pi  r_{+}^4 \sqrt{\frac{4 \eta ^2}{\beta ^2 r_{+}^2}+1} \left(6 -8 \pi  \left(\beta ^2 r_{+}^2 \left(\sqrt{\frac{4 \eta ^2}{\beta ^2 r_{+}^2}+1}-1\right)-2 \eta
   ^2\right)-\Lambda  r_{+}^2\right)}{8 \pi  \left(\beta ^2 r_{+}^2 \left(\sqrt{\frac{4 \eta ^2}{\beta ^2 r_{+}^2}+1}-1\right)-2 \eta ^2 \sqrt{\frac{4 \eta
   ^2}{\beta ^2 r_{+}^2}+1}\right)-\left(\Lambda  r_{+}^2+6\right) \sqrt{\frac{4 \eta ^2}{\beta ^2 r_{+}^2}+1}}.\label{eq:C2}
\end{equation}


As can be seen also in this $D=6$ case, the temperature and phase diagram in the upper panels of Fig.~\ref{fig:comparing6Tvrn6Cvr} behave exactly like ordinary Tangherlini-AdS black hole. The temperature will have a local minimmum and the specific heat imply two different state (unstable thermal AdS state and stable large black hole state) in the phase diagram. These will also be the case when we discuss $D=8$ black holes with DBI global monopole in the later subsection.

\subsubsection{$D=8$ case}


\begin{figure}[tp]
\begin{tabular}{cc}
  \includegraphics[width=8cm]{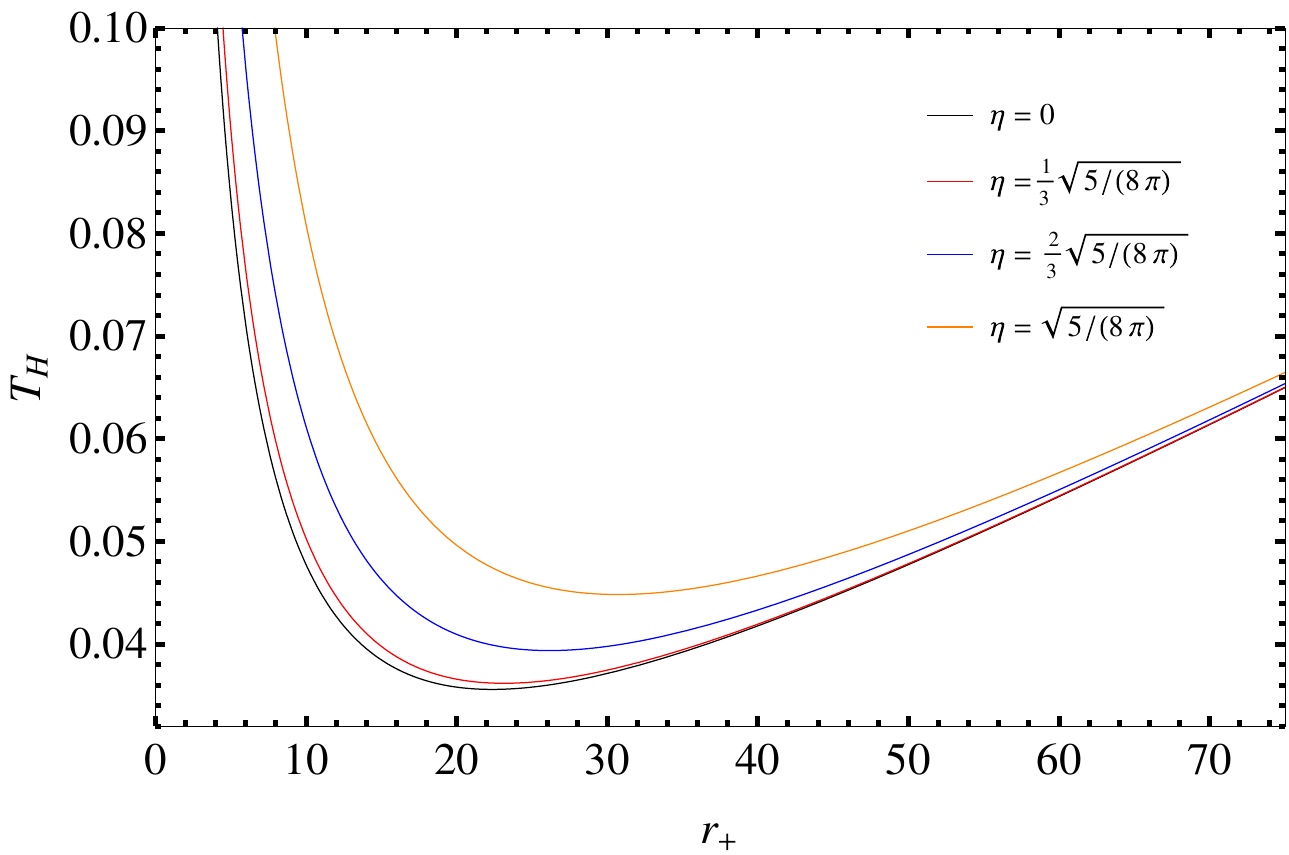} &   \includegraphics[width=8cm]{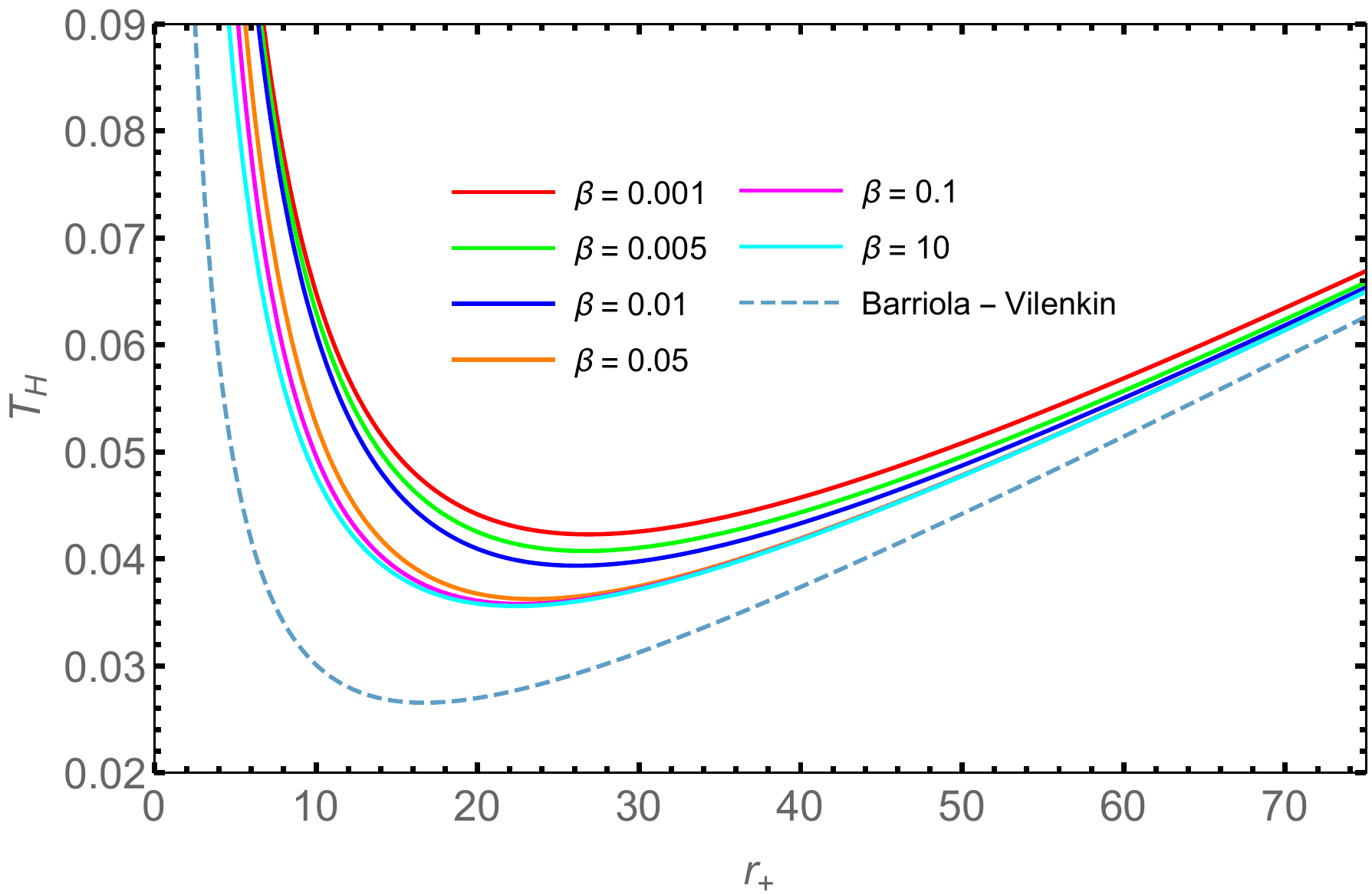} \\
(a) & (b) \\[6pt]
  \includegraphics[width=8cm]{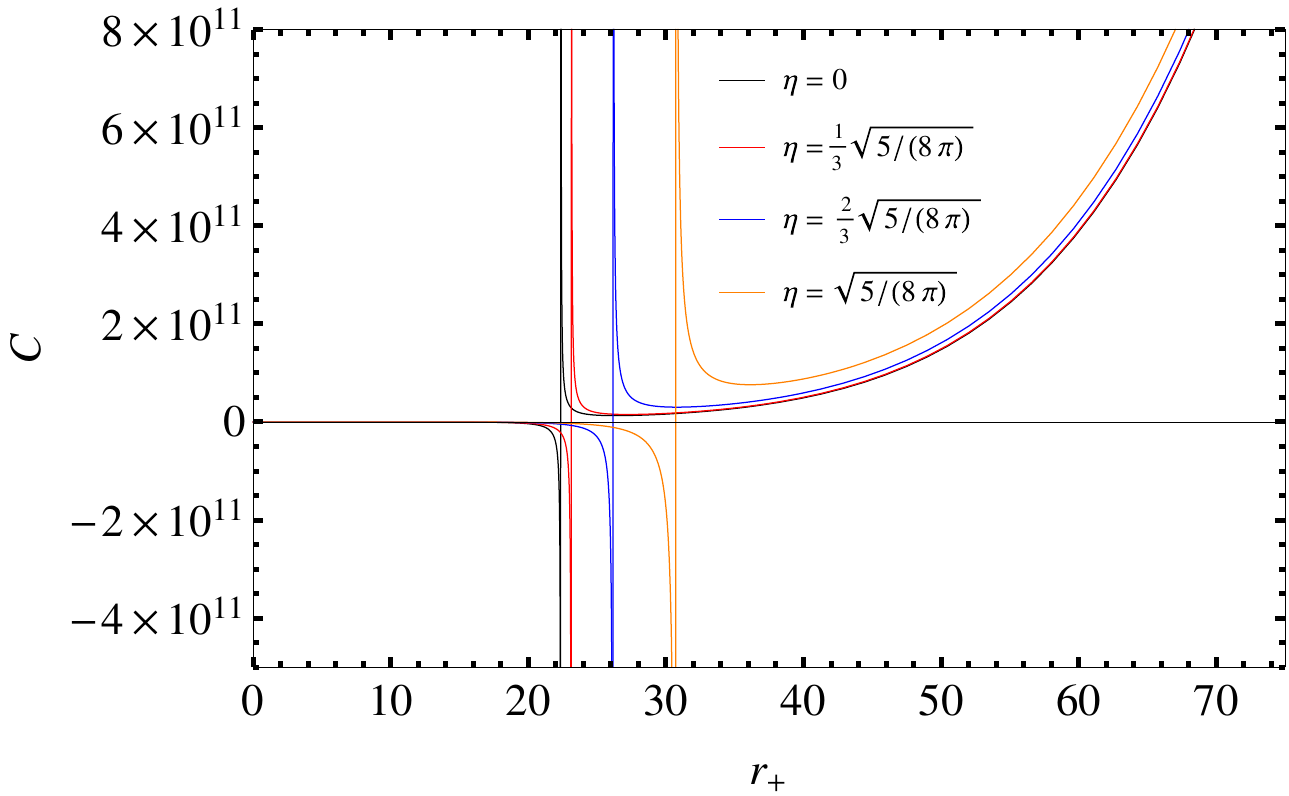} &   \includegraphics[width=8cm]{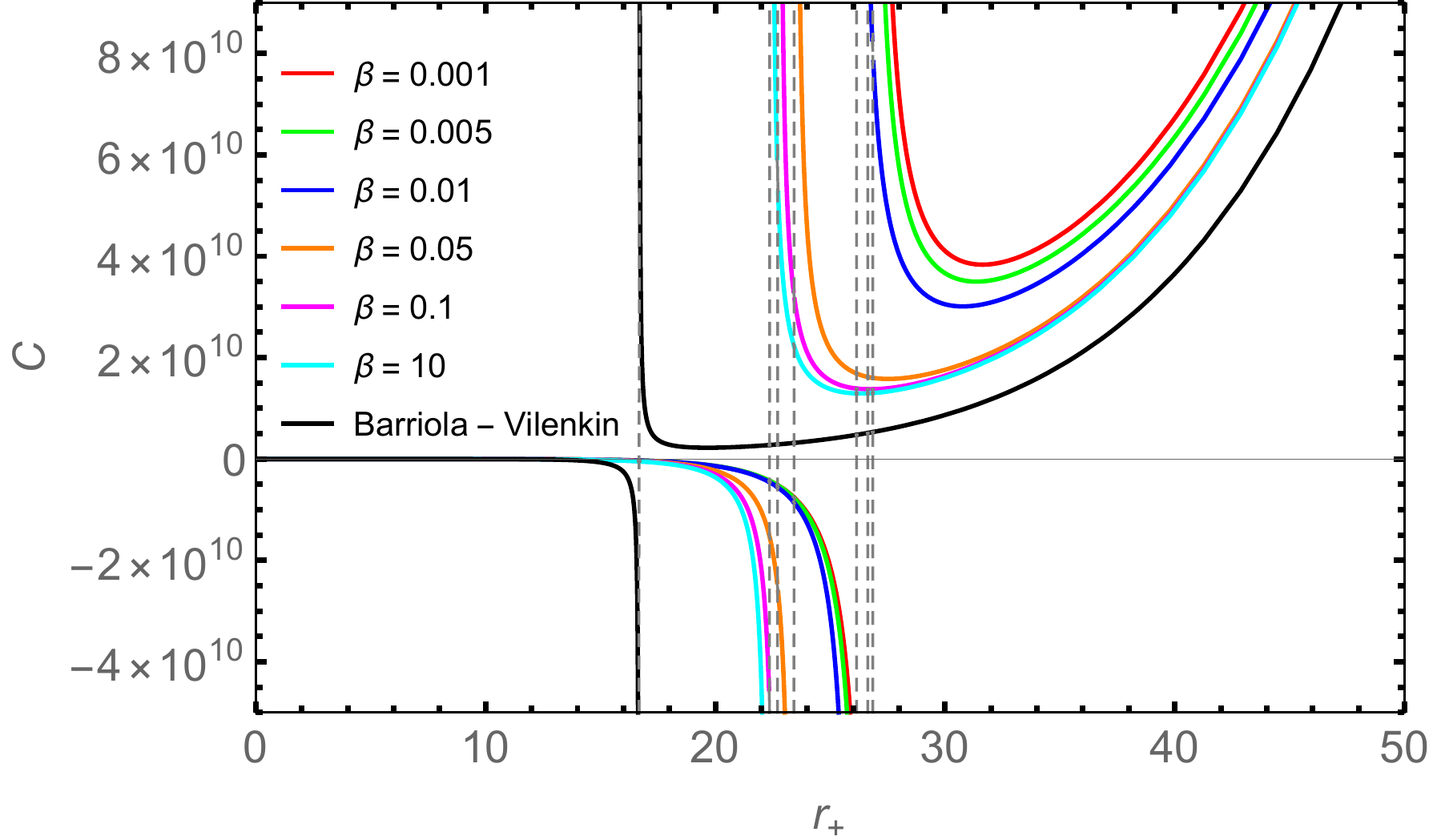} \\
(c) & (d) \\[6pt]
\end{tabular}
\caption{Hawking temperature as a function of event horizon, for $D=8$ AdS black holes with the value of $(a)\beta=0.01$ and $(b)\eta=0.297$, and also its specific heat as a function of event horizon, with $(c)\beta=0.01$ and $(d)\eta=0.297$.}
\label{fig:comparing8Tvrn8Cvr}
\end{figure}

From equation (\ref{eq:TH0}), the temperature of black hole with $D=8$ is given by
\begin{equation}
T_H = \frac{15 + \pi  \left(24 \eta ^2-8 \beta ^2 r_{+}^2 \left(\sqrt{\frac{6 \eta
   ^2}{\beta ^2 r_{+}^2}+1}-1\right)\right)-\Lambda  r_{+}^2}{12 \pi  r_{+}}.
\end{equation}
The behaviour of black hole temperature with various value of $\eta$ is given in Fig.~\ref{fig:comparing8Tvrn8Cvr}. The critical value is $\eta_{crit}=\sqrt{5/(8\pi)}$.


The entropy for $D=8$ black hole with DBI global monopole is $S = \left(\pi r^{6}/3\right)$. The specific heat for $D=8$ is given by
\begin{equation}
C_H = \frac{2 \pi  r_{+}^6 \sqrt{\frac{6 \eta ^2}{\beta ^2 r_{+}^2}+1} \left(15 -8 \pi 
   \left(\beta ^2 r_{+}^2 \left(\sqrt{\frac{6 \eta ^2}{\beta ^2
   r_{+}^2}+1}-1\right)-3 \eta ^2\right)-\Lambda  r_{+}^2\right)}{8 \pi 
   \left(\beta ^2 r_{+}^2 \left(\sqrt{\frac{6 \eta ^2}{\beta ^2
   r_{+}^2}+1}-1\right)-3 \eta ^2 \sqrt{\frac{6 \eta ^2}{\beta ^2
   r_{+}^2}+1}\right)-\left(\Lambda  r_{+}^2+15\right) \sqrt{\frac{6 \eta
   ^2}{\beta ^2 r_{+}^2}+1}}.
\end{equation}

The variation of $\eta$ and $\beta$ on the free energy of black holes in $D=8$ is similar with black holes in $D=4$, in which the on-shell free energy decreases as we increase the value of $\eta$ and the black holes will have lower on-shell free energy than the weakly-coupled case. As $\beta$ grows the phase transition can happen for larger radii.




\subsection{Thermodynamical behaviour of black holes in odd dimensions}

\begin{figure}[tp]
\begin{tabular}{cc}
  \includegraphics[width=8cm]{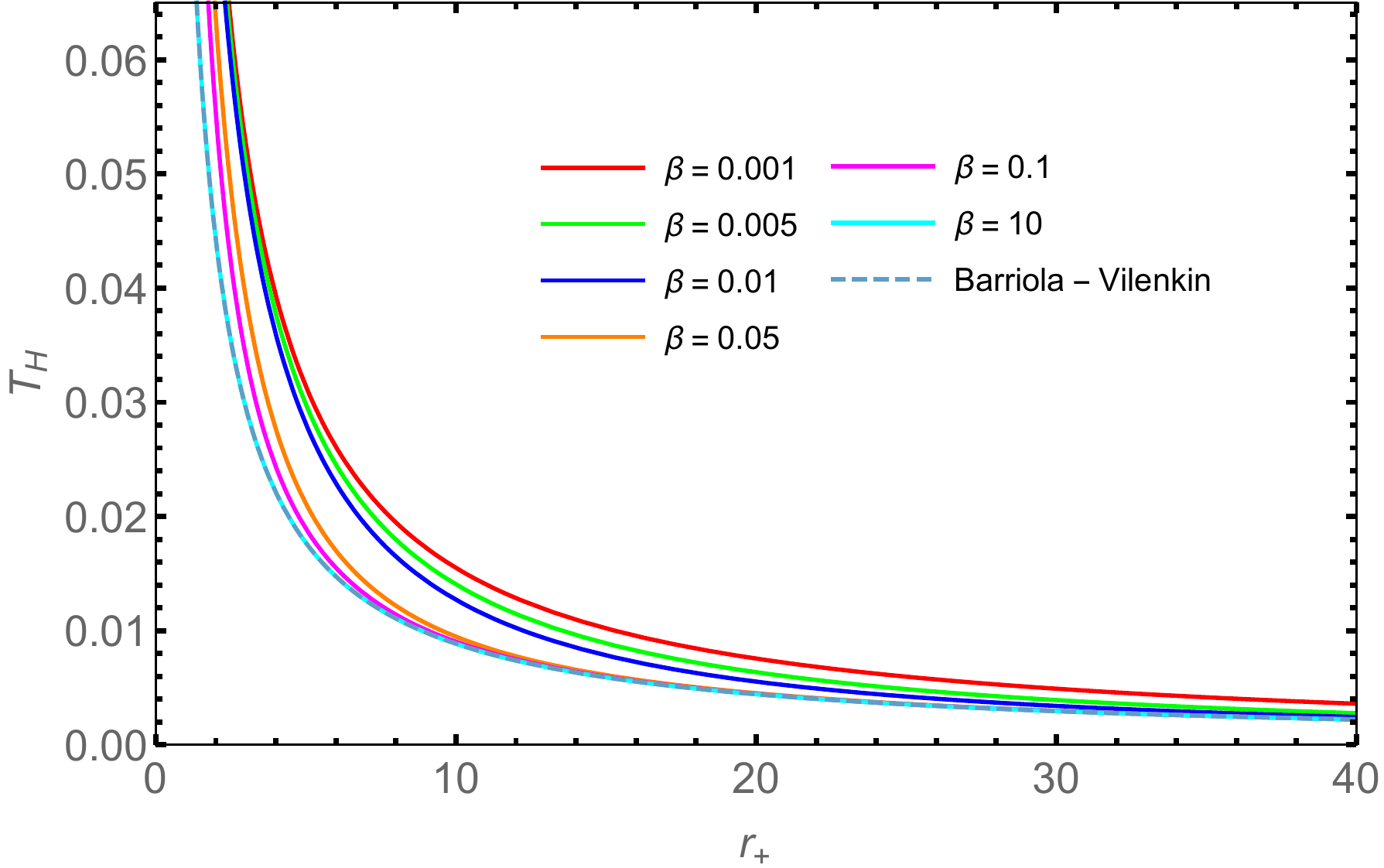} &   \includegraphics[width=8cm]{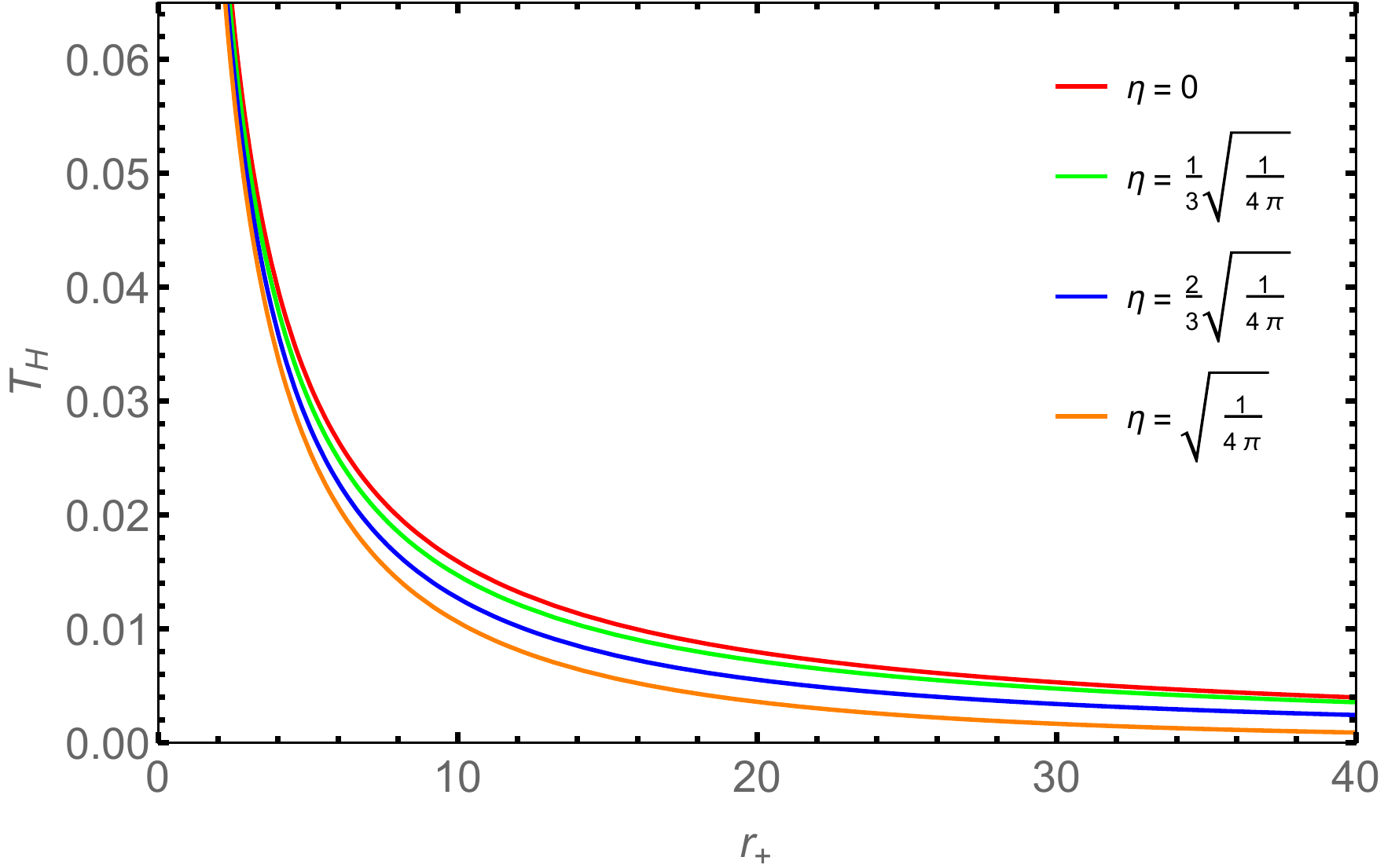} \\
(a) & (b) \\[6pt]
  \includegraphics[width=8cm]{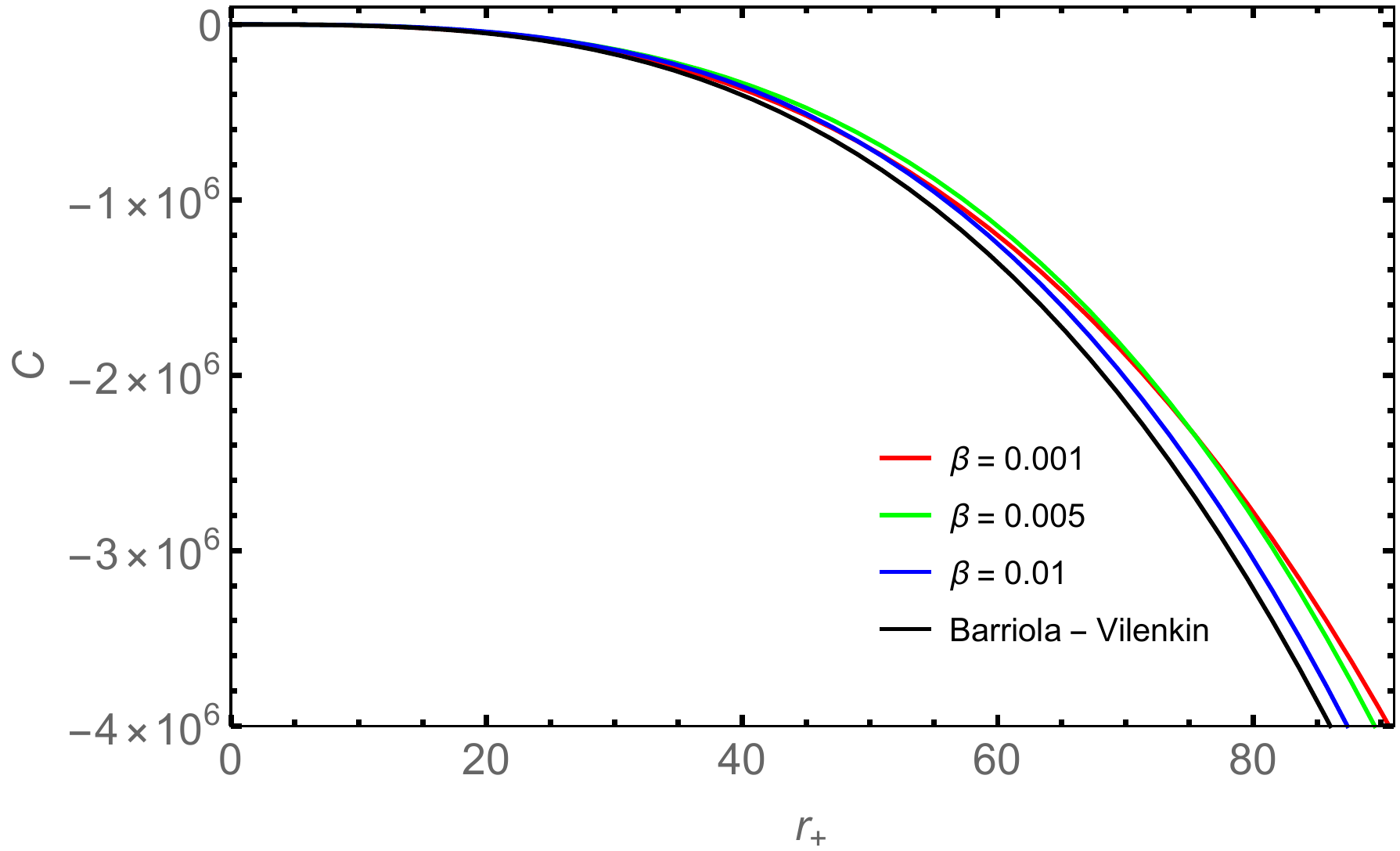} &   \includegraphics[width=8cm]{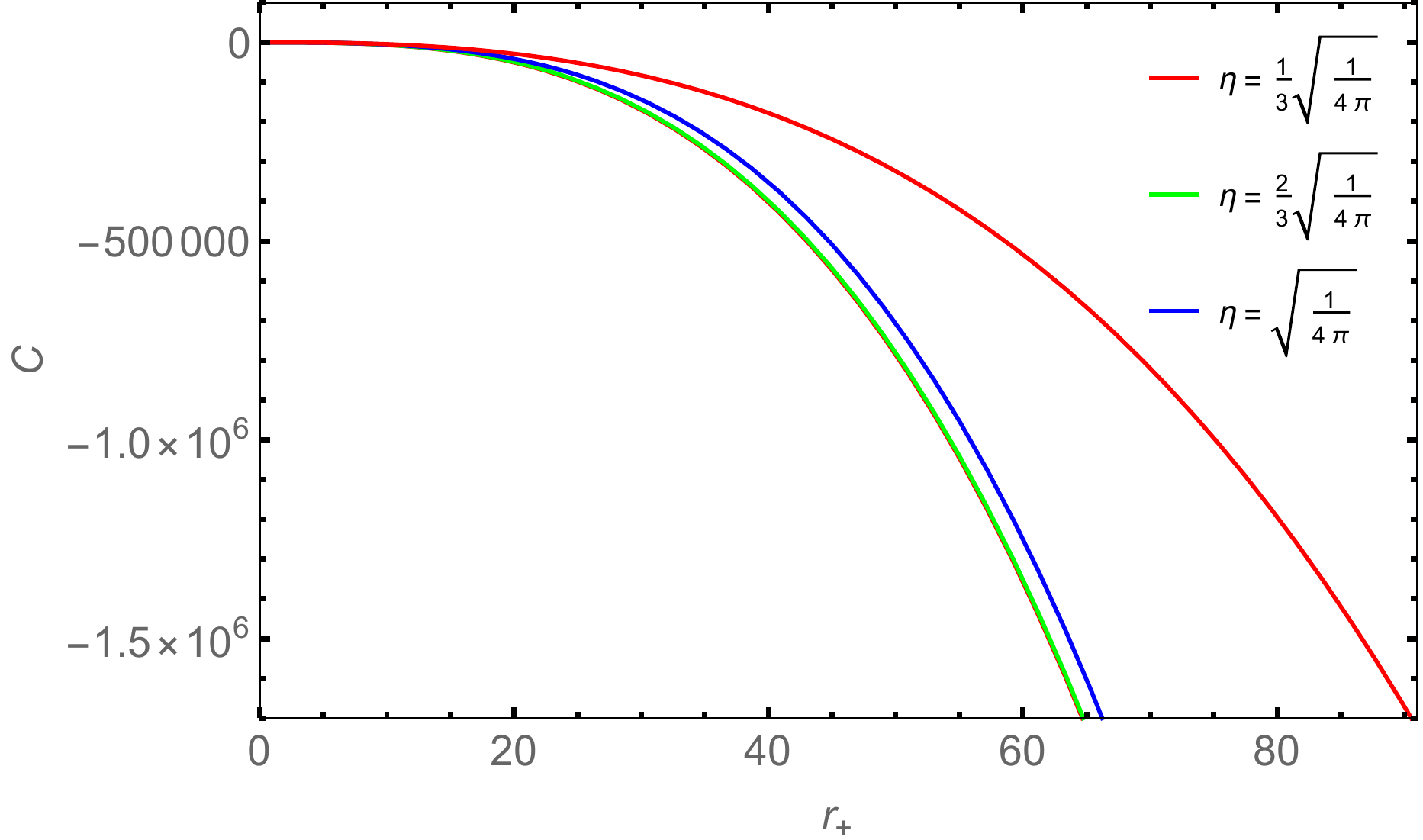} \\
(c) & (d) \\[6pt]
\end{tabular}
\caption{Hawking temperature as a function of event horizon, for $D=5$ asymptotically flat black holes with the value of $(a)\eta=0.188$ and $(b)\beta=0.01$, and also its specific heat as a function of event horizon, with $(c)\eta=0.188$ and $(d)\beta=0.01$.}
\label{fig:comparingfn5}
\end{figure}

As explained in the previous section, we can obtain an exact solution for black holes with odd dimensions by integrating the Einstein equation with a specific $D$. To investigate the thermodynamical behavior of black holes in odd dimensions, we shall analyze the Hawking temperature and specific heat. For the sake of concreteness, let us analyze $5d$ black holes whose solutions are given in~\eqref{odd2}.

The Hawking temperature for this dimension is given by
\begin{equation}
T_{H} = \frac{3-r^2 \left(\Lambda +8 \pi  \beta ^2 \left(\sqrt{\frac{3 \eta ^2}{\beta ^2 r^2}+1}-1\right)\right)}{6 \pi  r},
\end{equation}
which, quite surprisingly, does not depend on the logarithmic nature of the solution. Further, we can easily calculate the specific heat of the black holes. The equation for the specific heat is as follows
\begin{equation}
C_{H} = -\frac{2 \pi  r^3 \sqrt{\frac{3 \eta ^2}{\beta ^2 r^2}+1} \left(r^2 \left(\Lambda +8 \pi  \beta ^2 \left(\sqrt{\frac{3 \eta ^2}{\beta ^2 r^2}+1}-1\right)\right)-3\right)}{8 \pi 
   \beta ^2 r^2 \left(\sqrt{\frac{3 \eta ^2}{\beta ^2 r^2}+1}-1\right)-\left(\Lambda  r^2+3\right) \sqrt{\frac{3 \eta ^2}{\beta ^2 r^2}+1}}.
\end{equation}

\begin{figure}[tp]
\begin{tabular}{cc}
  \includegraphics[width=8cm]{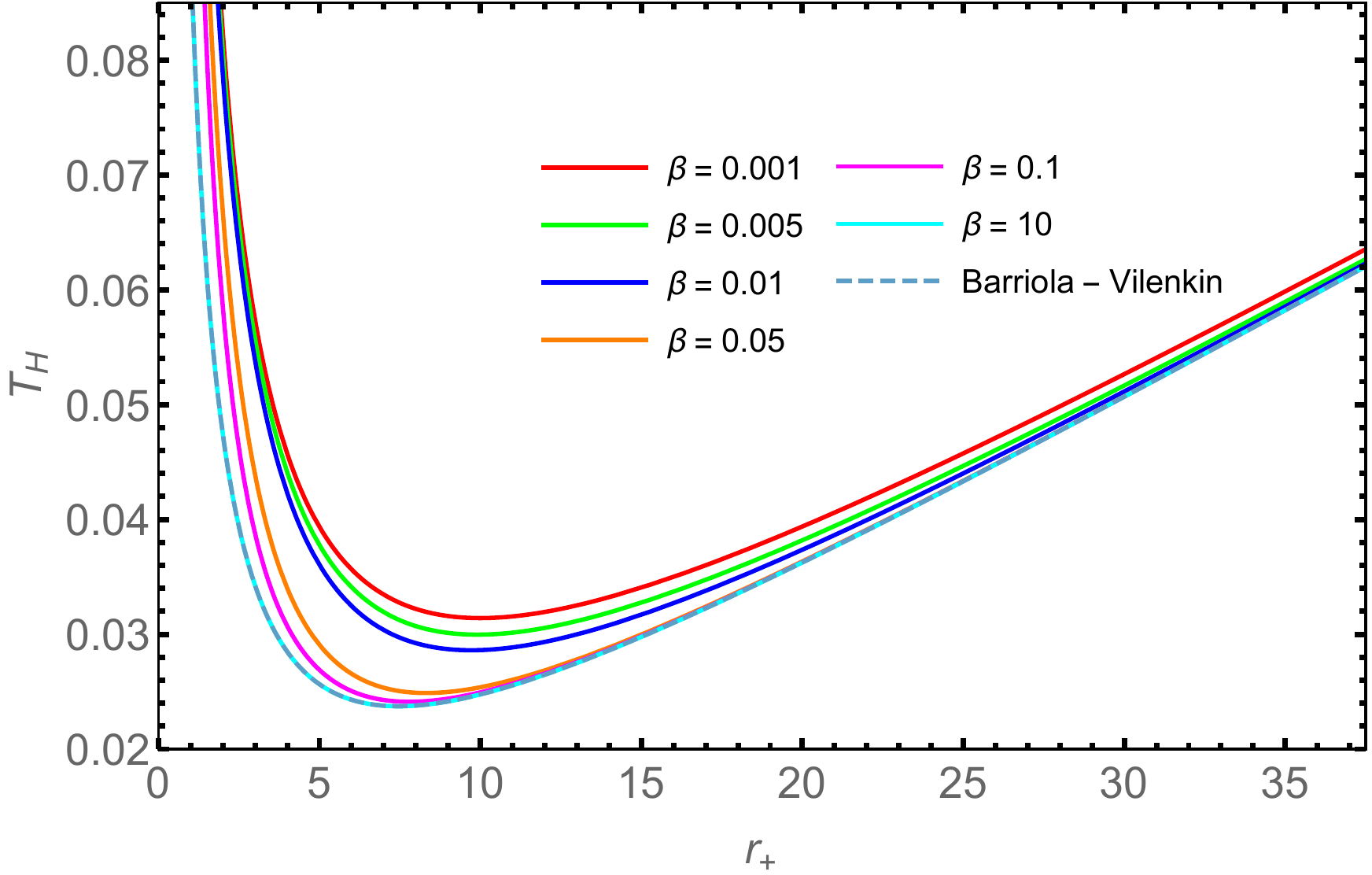} &   \includegraphics[width=8cm]{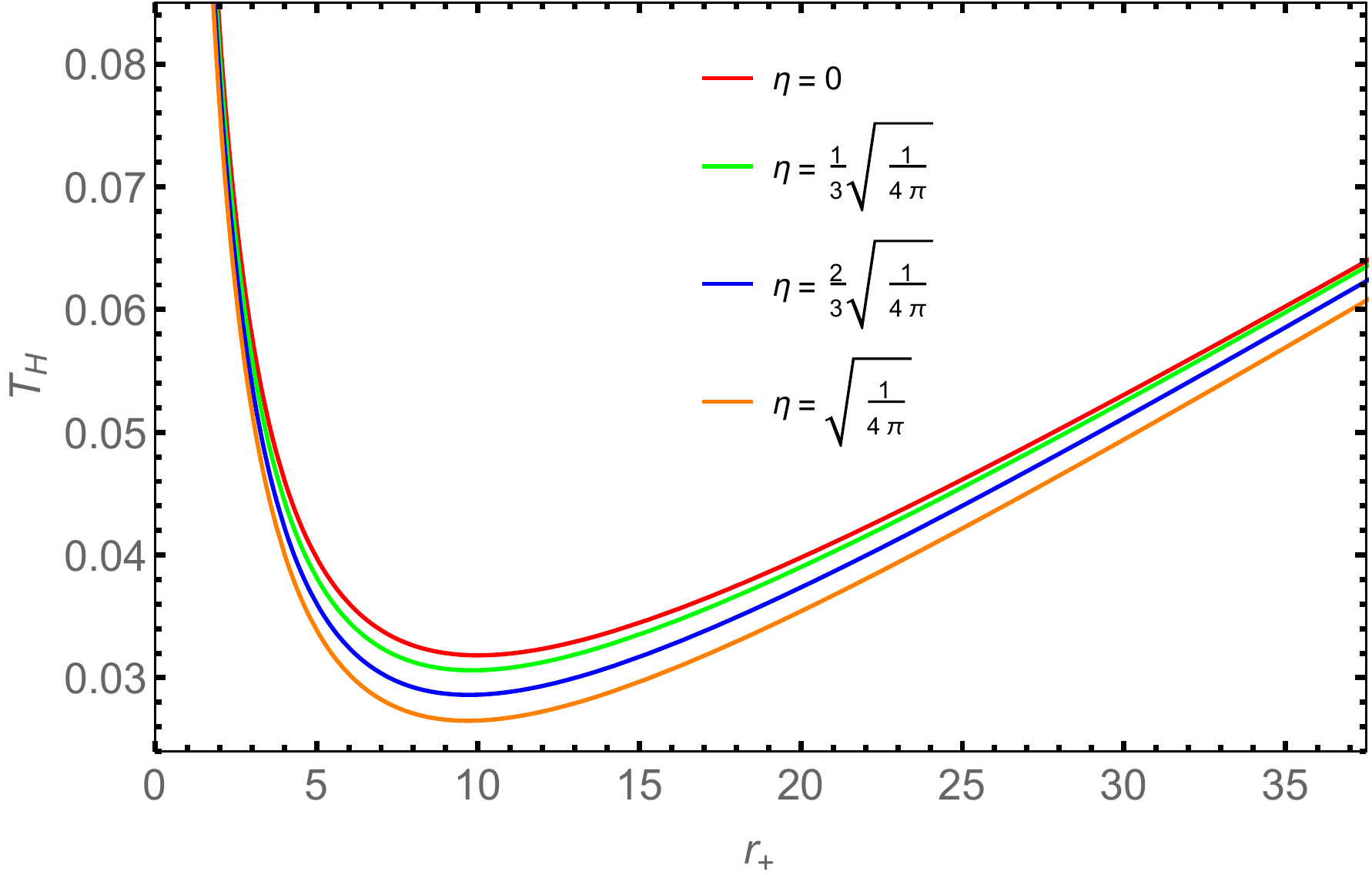} \\
(a) & (b) \\[6pt]
  \includegraphics[width=8cm]{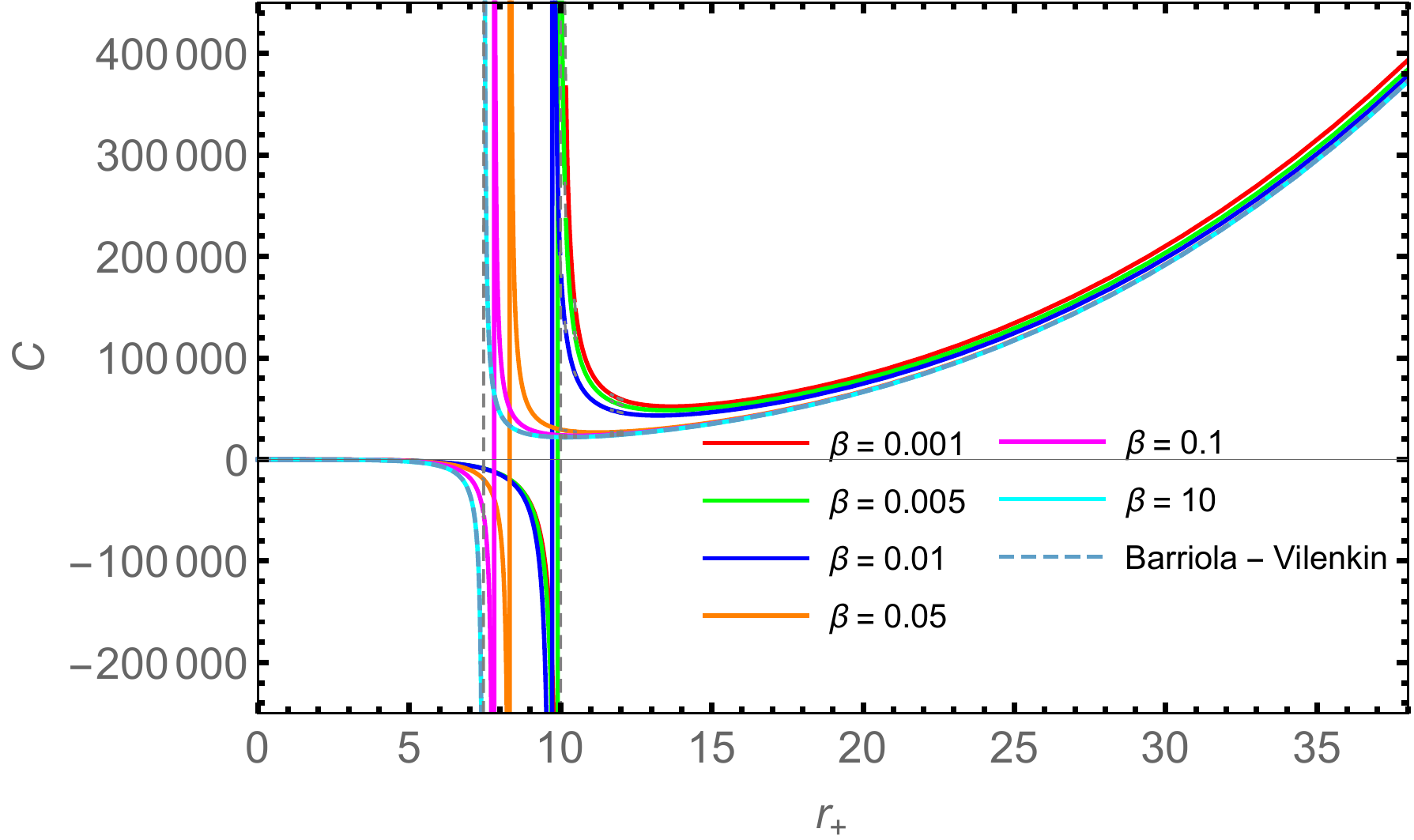} &   \includegraphics[width=8cm]{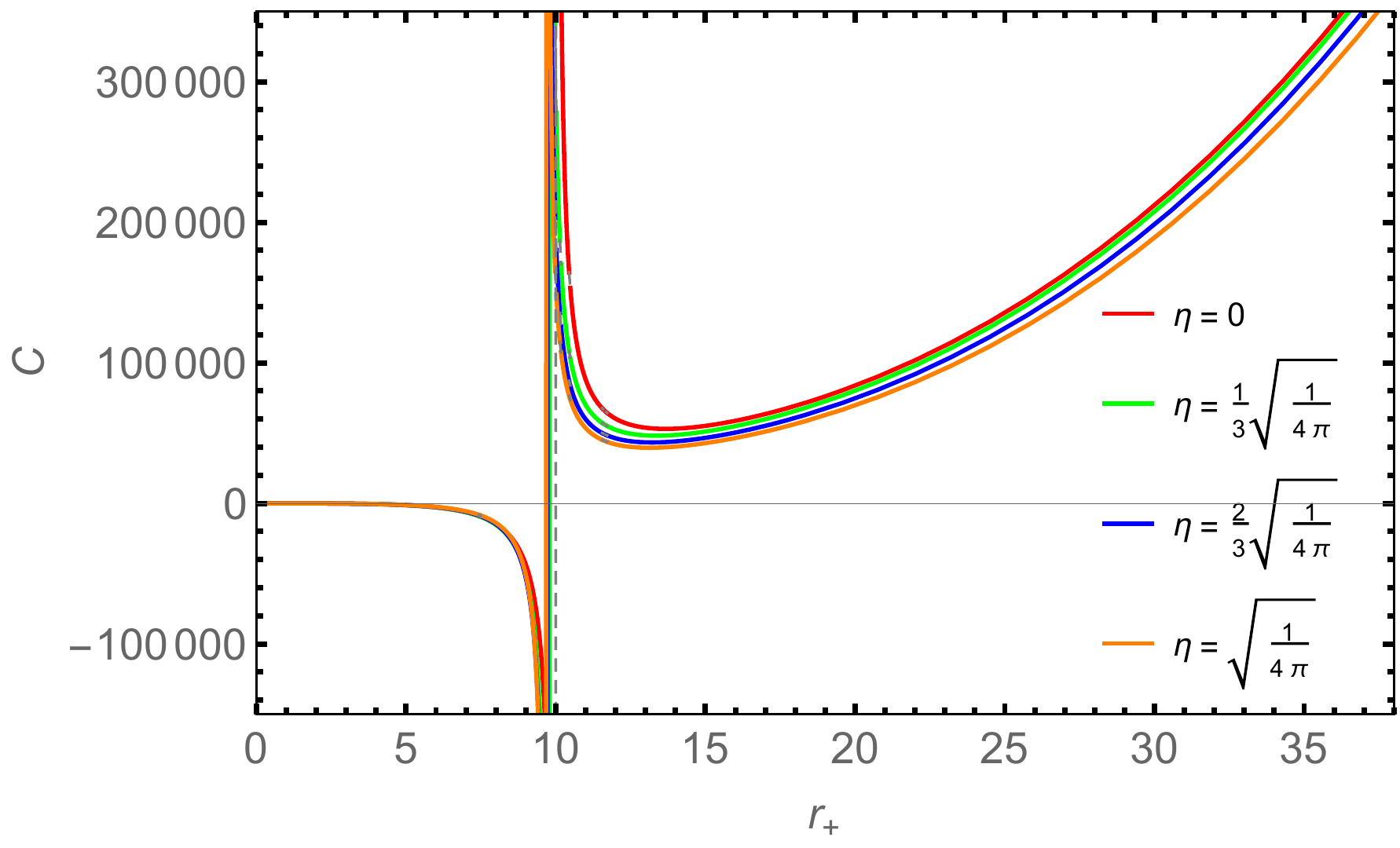} \\
(c) & (d) \\[6pt]
\end{tabular}
\caption{Hawking temperature as a function of event horizon, for $D=5$ AdS black holes with the value of $(a)\eta=0.188$ and $(b)\beta=0.01$, and also its specific heat as a function of event horizon, with $(c)\eta=0.188$ and $(d)\beta=0.01$.}
\label{fig:comparingadsn5}
\end{figure}

We concur that black holes in $D=5$ are thermodynamically feasible since there exists a set of thermal bodies properties that can be used to describe the said black holes, such as Hawking temperature and specific heat. The main thermodynamic structure of asymptotically flat and AdS black holes shown in even dimension cases are still evident in $D=5$, yet interestingly there are several new features shown here. It looks like the behavior of both asymptotically flat and AdS black holes with weakly coupled $\beta$ is indistinguishable with BV black holes. This behavior is shown in the Hawking temperature and the specific heat diagram, Figs.\ref{fig:comparingfn5}-\ref{fig:comparingadsn5}. It also appears that varying the values of $\eta$ has the opposite effect on the behaviour of $D=5$ black holes. In $D=5$ case, as the $\eta$ approaches the critical angle, its critical radius will also be smaller. Hence, we can say that black holes that have bigger $\eta$ will radiate faster compared to its counterpart with lesser value of $\eta$.

\section{Factorization solutions}\label{sec:factor}

Last but not least, in this section we shall discuss another type of solution which are non-black hole but instead a direct product of two-spaces of constant curvature. To this aim, let us choose an ansatz\footnote{This ansatz, and the solutions thereof, can be viewed as the nonlinear generalization to the spontaneous compactification due to scalar proposed in~\cite{GellMann:1984sj}.} $C(r)=C=constant$. This changes the components of the Einstein tensor 
\begin{eqnarray}
G^0_0=G_r^r={(D-2)(D-3)\over2C^2},~~ G^\theta_\theta=-{1\over B^2}{A''\over A}+{1\over B^2}{A'B'\over AB}+{(D-4)(D-3)\over2C^2}.
\end{eqnarray}
From $G_0^0$ we have
\begin{eqnarray}
{(D-2)(D-3)\over 2C^2}-\Lambda+8\pi G\beta^2\left(1-\sqrt{1+{(D-2)\eta^2\over \beta^2C^2}}\right)=0.
\label{pers02}
\end{eqnarray}
This gives us three and two possible value of $C^2$ when $\Lambda\neq0$  and $\Lambda=0$, repectively. 
When $\Lambda=0$, this gives us
\begin{equation}
C^2=\frac{(D-3)^2 (D-2)}{32\pi G\beta^2 (3-D-8\pi G\eta^2)},\label{pers05}
\end{equation}
thus in order to obtain a real compactification radius, $C^2>0$, we should have $D>3$ and $\eta^2<\eta^2_{crit}$; no additional constraint we can infer from this condition. 

For $\Lambda\neq 0$, on the other hand, 
\begin{eqnarray}
C^2_\pm &=& \left[ \beta^2 8\pi G(8\pi G\eta^2-D+3)+(D-3)\mp \beta 8\pi G \sqrt{\beta^2(8\pi G\eta^2-D+3)^2 +2(D-3)\eta^2\Lambda} \right]\nonumber
\\&&\times\frac{(D-2)}{2\Lambda(\Lambda-16\pi G\beta^2)}.\label{pers06}
\end{eqnarray}
There is no loss of generality should we $C^2_+$. This solution is valid ($C^2_+>0$) only for $\Lambda>16\pi G\beta^2$ or $\Lambda<0$, and under this range
\begin{equation}
\eta^2\geq \frac{D-3}{8\pi G}-\frac{(D-3)\Lambda}{\beta^2(8\pi G)^2} +\sqrt{\frac{(D-3)\Lambda(\Lambda-16\pi G\beta^2)}{\beta^4(8\pi G)^4}}\equiv\eta^2_{crit2}.
\end{equation}
Note that $\eta^2_{crit2}>\eta^2_{crit}$ for $\Lambda<0$, and $\eta^2_{crit2}<\eta^2_{crit}$ for for $\Lambda>16\pi G\beta^2$.

To obtain the metric solutions we employ $G^\theta_\theta-G^r_r$ to get
\begin{equation}
B^{-2}\left({A'B'\over AB}-{A''\over A}\right)=\pm\omega^2, \label{pers03}
\end{equation}
where we define a real $\omega$ with
\begin{equation}
\pm\omega^2\equiv{D-3\over2C^2}-{4\pi G\eta^2/C^2\over \sqrt{1-{(D-2)\eta^2\over\beta^2 C^2}}}.
\end{equation}
In general, $\omega^2$ can be positive, negative, or zero. Now, we still have gauge freedom in the metric. To fix it, we can take set $B=1$, and Eq. (\ref{pers03}) yields
\begin{eqnarray}
ds^2=
\begin{cases}
\frac{1}{\omega^2}(\sin^2 \chi ~dt^2 - d\chi^2) -C^2 d\Omega_{D-2}^2, & \text{for} \omega^2>0,\\
dt^2 -dr^2 -C^2 d\Omega_{D-2}^2, & \text{for } \omega=0,\\
\frac{1}{\omega^2}(\sinh^2 \chi ~dt^2 - d\chi^2) -C^2 d\Omega_{D-2}^2, & \text{for} \omega^2<0,
\end{cases}\label{eq:B1}
\end{eqnarray}
with $\chi\equiv\omega r$. Another gauge we can employ is $B\equiv A^{-1}$. In this form, the solutions are as follows 
\begin{eqnarray}
ds^2=
\begin{cases}
(1 -\omega^2 r^2)~dt^2 -(1 -\omega^2 r^2)^{-1}dr^2 - C^2 d\Omega_{D-2}^2, & \text{for} \omega^2>0,\\
dt^2 -dr^2 - C^2 d\Omega_{D-2}^2, & \text{for } \omega=0,\\
(1 +\omega^2 r^2)~dt^2 -(1 +\omega^2 r^2)^{-1} dr^2 - C^2 d\Omega_{D-2}^2, & \text{for} \omega^2<0.
\end{cases}\label{eq:BA1}
\end{eqnarray}
The readers can easily verify that solutions~\eqref{eq:B1} and \eqref{eq:BA1} are nothing but the same spacetimes written in different gauge. They are direct products of two-spaces of constant curvature: Nariai ($dS_{2}\times S^{D-2}$)~\cite{Nariai} for $\omega^2>0$, Plebanski-Hacyan ($M_{2}\times S^{D-2}$)~\cite{Plebanski} for $\omega^2=0$, and Bertotti-Robinson ($AdS_{2}\times S^{D-2}$)~\cite{Bertotti:1959pf, Robinson:1959ev} for $\omega^2<0$. 

In short,there are three possible higher-dimensional spaces (depending on the sign of $\Lambda$), $X_D$, that can be factorized in three possible channels, $Y_2\times S^{D-2}$. Knowing these we should determine which channels are allowed, at least classically. This can be done by checking whether the condition satisfying $\omega^2$ simultaneously also holds for $C^2>0$. Solving the polynomial equations $\omega^2>0$, $\omega^2=0$, or $\omega^2<0$. The results are shown in Table (\ref{table:tangDBI}). These tell us that the possible channels of factorization are
\begin{eqnarray}
	dS_D &\longrightarrow& dS_2 \times S^{D-2},\\
	M_D &\longrightarrow& dS_2 \times S^{D-2},\label{eq:channelunstable}\\ 
	AdS_D &\longrightarrow&
	\begin{cases}
		dS_2\times S^{D-2},\\
		M_2\times S^{D-2}.
	\end{cases}
\end{eqnarray}

\begin{table}[h]
	\caption{Conditions for DBI monopole compactification in $D$ dimensions.}
	\begin{tabular}{cccc}
		\hline \rule[-2ex]{0pt}{5.5ex}  & \,\,$dS_2\times S^{D-2}$\,\, & \,\,$M_2\times S^{D-2}$\,\, & \,\,$AdS_2\times S^{D-2}$ \\ 
		\hline \hline \rule[-2ex]{0pt}{5.5ex} $\Lambda>16\pi G\beta^2$\ \ \; & $\frac{(D-3)\beta^2}{2(8\pi G\beta^2-\Lambda)}\neq\eta^2\geq\eta^2_{crit2}$ \ \  &\ \  forbidden \ \ \ \ &\ \  forbidden \ \  \ \ \\ 
		\hline \rule[-2ex]{0pt}{5.5ex} $\Lambda=0$\ \ \; & $\eta^2>\eta^2_{crit}\ \ $ & forbidden\ \ \ \  &\ \  forbidden\ \ \ \ \\ 
		\hline \rule[-2ex]{0pt}{5.5ex} $\Lambda<0$\ \ \; & \;$\eta^2>\eta^2_{crit2}\ \ $\; & \; $\eta^2=\eta^2_{crit2}\ \ \ \ $ \; & \; forbidden\ \ \ \ \; \\ 
		\hline 
	\end{tabular} 
	\label{table:tangDBI}
\end{table}
It is unfortunate, from Table~\eqref{table:tangDBI}, that factorization $AdS_D\rightarrow AdS_s\times S^{D-2}$ is forbidden in our theory, since $AdS_2\times Y$ (with $Y$ any manifold depending on the context) appears quite generally as the near-horizon limit of extremal black holes geometry; {\it e.g.,} near-horizon limit of extremal Reissnerr-Nordstrom black hole is topologically $AdS_2\times S^2$.  

\section{Conclusions}\label{sec:conclusion}

There are three aims we wish to establish in this work. First, we confirm the equivalence of exact solutions of Einstein's equations with non-linear $\sigma$ model and approximate metric solutions outside a global defects. We generalize this equivalence to include the non-canonical $k$-form in the non-linear $\sigma$-model. The particular form we consider is the Dirac-Born-Infeld (DBI). We assume higher-dimensional spherical symmetry equipped with cosmological constant. This theory can be perceived as an extension of our previous result~\cite{Prasetyo:2015bga, Prasetyo:2017rij}. Our investigation shows such an equivalence, as has also been employed in~\cite{Olasagasti:2000gx, Tan:2017egu, Prasetyo:2017rij}. 

Second, by taking the hedgehog ansatz for the scalar field, we obtain exact solutions of gravitating $\sigma$-model which can be interpreted as metric outside global defects. Such metric suffers from deficit solid angle $\Delta\equiv\frac{8 \eta ^2 \pi  G}{D-3}$. The solutions exist in all dimensions, but can be expressed as polynomial functions only in even D. They can be perceived as a gravitational field outside a (DBI) global monopole, as in a generalization of Barriola-Vilenkin (BV) solutions~\cite{Barriola:1989hx}, as well as a black hole eating up such a monopole, depending on the strength of the mass. Here, we focus more on the solution's interpretation as black holes with DBI global hair. This can be regarded as a non-canonical and higher-dimensional generalization of the solutions studied in~\cite{Dadhich:1997mh}. due to the nonlinearity of the kinetic term, the scalar charge cannot be rescaled away; thus our solution describes a genuine (higher-dimensional) black hole with scalar hair. Unlike our previous results with quadratic kinetic term for the scalar field~\cite{Prasetyo:2017rij}, here the dS black holes only possess at most two horizons.The DBI nature makes the scalar charge depend on all even powers of $r$ (Eq.~\eqref{eq:expandbars}). However, due to the alternating sign in the series the metric is unable to have more than one horizons for the case of asymptotically-flat and AdS space. This results in the existence of only one extremal horizon for the dS case. We investigate such extremal conditions for several dimension. We also give a complete relation between the mass and the scalar charge for the extremal condition to happen.

Having discussed the solutions we extensively investigate their thermodynamics properties, and this is our last purpose of this work. We found that the DBI modification yields genuine effects on its temperature, thermodynamical stability, and phase transition. While the ordinary ($4d$ or higher-dimensional) BV black holes have minimum temperature for the maximum solid angle, we show that instead the DBI black holes do have the least temperature in its Tangherlini limit ($\Delta\rightarrow0$); the higher the scalar charge the more they radiate. An interesting result is that the entropy of our solution is the same as the entropy of Tangherlini black hole with global monopole, $S = \frac{2 \pi  r_{+}^{D-2}}{G (D-2)}$, independent of the deficit solid angle $\Delta$ as well as the DBI coupling constant $\beta$. It suggests that the entropy is not affected by the UV-modification of a theory. From the specific heat capacity $C_H$ we can infer the stability of the corresponding solutions. They are stable if $C_H>0$ and unstable otherwise. We show in Figs.~\ref{fig:comparingn4Tvrn4Cvr} and \ref{fig:comparing6Tvrn6Cvr} for AdS black holes\footnote{All asymptotically-flat black holes are unstable, as can be seen from the negativity of the corresponding specific heat, Figs.~\ref{fig:fn4Tvrfn4Cvr} and \ref{fig:betafnCvr}.} that as $\Delta$ or $\beta$ increases, the singularity of $C_H$ shifts to the right, indicating the possibility of (first-order) phase transition. This as well implies that as the deficit angle gets higher or $\beta$ gets stronger the number of black holes (in a network full of them with arbitrary radii) that are stable against decay to pure AdS state get smaller because it takes higher radius to become stable. 

What is left out to discuss here is the thermal properties of SdS black holes with global monopole. In this work we only focus on the asymptotically-flat and AdS solutions for the discussion on thermodynamics. A few problems arises once we analyze the thermodynamics of dS black holes, and mainly it boils down to the problem of determining the appropriate definition of temperature in dS black holes. As we know, Schwarzschild black holes in de Sitter spacetime have two types of horizon, the event horizon (EH) and the cosmological horizon (CH). These horizons would have different surface gravity, hence both have its own temperature. If we only use the definition of Hawking temperature ($T_{BH}$), there will be an apparent instability in the black holes horizons, since the temperature of EH would be higher than that of CH \cite{Urano:2009xn}. We need to formulate a new temperature definition that would accommodate both horizons. One of the most well-known method is the Bousso-Hawking normalization, in which the Killing vector is accordingly normalized to a constant related to its geodesic orbit \cite{Bousso:1996au}. There are also studies conducted using the so-called standard normalization by employing Hawking temperature ($T_H$) and Gibbons-Hawking temperature ($T_{GH}$) as the definition to describe the thermal properties of EH and CH \cite{Myung:2006tg,Myung:2007my}. Recently, a study has been done to compare different types of temperature definition, and it is shown that the Bousso-Hawking normalized temperature still works as the most adequate definition of SdS black holes \cite{Kanti:2017ubd}. The indication that the cosmological constant of our universe might be a positive one naturally increases the attention being put into studying the thermal stability of SdS black holes, and we refer to the literature mentioned above for extensive and rigorous reviews on this matter.

\section{Acknowledgements}

We thank Gema Darman and Reyhan Lambaga who helped us with stimulating discussions during the initial phase of this project. We also thank the anonymous referees for the comments and critics of the manuscript. This work is supported by the PITTA grant from Universitas Indonesia under the contract no.~656/UN2.R3.1/HKP.05.00/2017.


\end{document}